\documentclass[12pt,preprint]{aastex}
\usepackage{epsfig}

\shorttitle{Explosive hydrogen burning during X-ray bursts}
\shortauthors{Fisker, Schatz & Thielemann}
\begin{document}
\title{Explosive hydrogen burning during type I X-ray bursts}
\author{Jacob Lund Fisker}
\affil{Department of Physics and Joint Institute for Nuclear Astrophysics,
        University of Notre Dame,
        Notre Dame, IN 46566,
        USA}
\email{jfisker@nd.edu}
\author{Hendrik Schatz}
\affil{Department of Physics and Astronomy and National Superconducting Cyclotron Laboratory and Joint Institute for Nuclear Astrophysics, Michigan State University, East Lansing, MI 48824-2320}
\email{schatz@nscl.msu.edu}
\author{Friedrich-Karl Thielemann}
\affil{Department of Physics and Astronomy, University of Basel, Klingelbergstrasse 82, 4056 Basel, Switzerland}
\email{F-K.Thielemann@unibas.ch}
\begin{abstract}
Explosive hydrogen burning in type I X-ray bursts (XRBs) comprise charged particle reactions creating isotopes with masses up to $A\sim 100$. Since charged particle reactions in a stellar environment are very temperature sensitive, we use a realistic time-dependent general relativistic and self-consistent model of type I x-ray bursts to provide accurate values of the burst temperatures and densities. This allows a detailed and accurate time-dependent identification of the reaction flow from the surface layers through the convective region and the ignition region to the neutron star ocean. Using this, we determine the relative importance of specific nuclear reactions in the X-ray burst.
\end{abstract}


\section{Introduction}
Type I X-ray Bursts (XRBs) (see \cite{Bildsten98c,Strohmayer06} for reviews)
were first explained by \cite{Woosley76} who associated the XRBs with
thermonuclear runaways on the surface of neutron stars that accrete a mixture
of hydrogen and helium from a semi-detached low mass companion star
\citep{Joss77}. 

In Woosley \& Taam's \emph{thermonuclear flash model} the impact of the
accreting matter fully ionizes the matter and heats it to $1-2\times
10^8\,\textrm{K}$ explaining the persistently observed X-ray emissions.  The
accreted matter then undergoes gradual compression as it sinks as new matter
continuously piles on top of it.
Under these atmospheric conditions the electrons are degenerate, but the
nucleons are not. Therefore the matter is subject to a thin-shell instability
that triggers a thermonuclear runaway \citep{Hansen75}.  For accretion rates
below roughly one Eddington ($\dot{M}=1.12\times
10^{18}\textrm{g}\,\textrm{s}^{-1}$), the bottom layer of the newly accreted
matter becomes unstable after a few hours/days and burns explosively and gives
rise to a burst of X-rays as the atmosphere is heated to 1-1.5 GK.
The sudden release of nuclear binding energy heats the atmosphere rapidly and
increases the luminosity within a few seconds after which the luminosity decays
exponentially as the atmosphere cools again producing the observational
features of a type I X-ray burst \citep{Joss78b,Taam80}.  These bursts are the
most common thermonuclear explosions in the universe, so the indestructable and
repeatable LMXBs are useful for indirect conclusions about the behavior of
matter under extreme conditions. 

X-ray bursts have been explored theoretically by
\cite{Hanawa83,Fujimoto87b,Koike99} and stable burning has been explored by
\cite{Schatz99} using relatively simple hydrodynamical models to estimate the
burning conditions e.g.~a set of $(\rho,T,\vec{X})$, where $\rho$ is the
density, $T$ is the temperature, and $\vec{X}$ is a composition array
describing the fractional concentration of each isotope.  On the other hand
more realistic 1 dimensional models have been constructed by many groups
\citep{Joss78b,Taam79,Hanawa83,Wallace82,Ayasli82} but they suffered from
relatively simple nuclear reaction networks so only recently models have
successfully included both aspects \citep{Woosley04,Fisker04b,Fisker06}.

The relevant types of reaction sequences in XRBs have been discussed by
\cite{Wallace81,Champagne92,Wormer94,Herndl95,Schatz98,Schatz06}.  Important
are $(p,\gamma$)-, $(\alpha,\gamma$)-, $(p,\alpha$)-reaction rates as 
well as $\beta$-decay rates between the
valley of stability and the proton dripline. 
Reaction rates must be known up to the end-point of the $rp$-process \citep{Schatz01}. 

In the past serveral attempts have been made to identify critical reaction 
rates in X-ray bursts.
It is, however, not possible to directly test the astronomical number of
possible perturbations of the thousands of participating reaction rates.
\cite{Woosley04} changed groups of decay rates and narrowed the rates down to
several important candidates. \cite{Fisker04,Fisker04b,Fisker06} relied on
``inspired guesses'' and found individual important rates. Recently,
\cite{Amthor06} has used a one-zone model and individually varied a large
number of reaction rates with the intent of verifying ``one-zone''-candidates
with a multi-zone model.  Using Monte Carlo methods, \cite{Roberts06} varied
random groups of reaction rates and similarly identified the most significant
candidates for later verification with multi-zone models. 

In this paper, we use such a full 1D X-ray burst model with a complete nuclear
reaction network to answer one of the fundamental questions in this field: what
are the nuclear reactions that power X-ray bursts?  While previous studies have
used simplified models to delineate basic types of reaction sequences we can
now go the next step and describe the actual nuclear reaction sequences that
occur as a function of time and depth during a typical X-ray burst. 
Because temperature, density, and initial composition vary greatly as a
function of depth there is no single reaction flow, but a range of very
different sequences that influence each other.  Identifying the nuclear
reactions that take place in X-ray bursts is a prerequiste for understanding
X-ray burst light curve features in terms of the underlying nuclear physics and
for determining the nuclear physics uncertainties in X-ray burst model
predictions of lightcurves and other observables.  It is also essential to
guide experimental and theoretical efforts to address these uncertainties in
the future. 

Cross sections have typically been predicted by global models which in most
cases have been fitted to stable nuclei and extrapolated to proton-rich nuclei.
In many cases cross sections have also been predicted by nuclear shell model
calculations.  However, with upgrades of existing experimental facilities and
new facilities many of these reactions are now in range of experiments
\citep{Kaeppeler98,Wiescher01,Wiescher01b,Schatz02}.

With a better understanding of the nuclear physics it will also become
possible to address potential issues beyond the 1D approximation, such as the
interplay of lateral flame propagation and nuclear energy release timescales
that can affect the modeling of burst rise times.

\section{The 1D multi-zone computational burst model}\label{sec:ourmodel}
In this paper we compute and describe one XRB model using the parameters $M=1.4 M_\odot$, $R=11\textrm{km}$ and a proper global accretion rate of $\dot{M}=1\cdot 10^{17}\textrm{g}/\textrm{s}$. This choice yields a H/He-ignited XRB corresponding to case (1) of \cite{Fujimoto81} where the observational data depends more on the $rp$-process than is the case for a pure He-ignited XRB (case (2) of \cite{Fujimoto81}). 
The Newtonian accretion rate (observable at infinity) of \cite{Schatz01,Schatz01b} and model \emph{zM} of \cite{Woosley04} was $\dot{M}=5\cdot 10^{16}\,\textrm{g}\,\textrm{s}^{-1}$, so the burst behavior of our model is expected to be similar but have a slightly higher hydrogen content and a slightly lower helium content at the point of ignition resulting in a longer rise time.

This model is calculated using a general relativistic type I X-ray burst simulation code that is described in more detail in \cite{Fisker06}. The code couples the general relativistic hydrodynamics code, \verb+AGILE+ \citep{Liebendoerfer02}, with the nuclear reaction network solver of \cite{Hix99}. 
The code includes radiative, conductive, and convective heat transport as described in \cite{Thorne77} and uses an arbitrarily relativistic and arbitrarily degenerate equation of state. 
We calculate the radiative opacities due to Thomson scattering and free-free absorption using the analytic formulations of \cite{Schatz99}. 
We use the same conductivity formulations for electron scattering on electrons, ions, phonons, and impurities as \cite{Brown00}. 

\verb+AGILE+ solves the general relativistic equations in a spherically symmetric geometry on a comoving grid. 
The computational domain covers about 7 pressure scale heights and is discretized into 129 log-ratioed grid zones with a column density\footnote{\label{footnote:relativistic-y} The relativistic column density is mass of a column above an area: $y\equiv\int_{R-r}^R\rho {dr \over \Gamma}$ where $\Gamma=\sqrt{1-2GM/Rc^2}$, so $P\simeq gy$, where $R$ is the neutron star radius, $M$ is the neutron star mass, $\rho$ is the density, $P$ is the pressure, and $g=GM/\Gamma R^2$ is the surface acceleration of gravity.} ranging from $y=1.2\times 10^6$ g cm$^{-2}$ to $y=3.9\times 10^9$ g cm$^{-2}$. 
The computational domain is bounded by a realistic core boundary interface \citep{Brown03,Brown04} and a relativistically corrected grey atmosphere \citep{Thorne77,Cox04} which is integrated numerically out to $P_{surf}=10^{18} \textrm{g}\,\textrm{cm}^{-2}$ using a 4th order Runge-Kutta method for greater accuracy \citep{Fisker06}.

The rp-process is naturally limited once it reaches the $A\sim 104$ region because neutron deficient nuclei in this mass range become $\alpha$-unbound. This terminates the reaction flow via ($\gamma$,$\alpha$) reactions and forms a SnSbTe cycle \citep{Schatz01}. 
This determines the maximum network size that is needed, unless the alpha unbound nuclei can be circumvented in multiple proton exposures.
\cite{Schatz01} demonstrated that the $A\sim 104$ endpoint can be reached if burst peak temperatures
and hydrogen concentration at ignition are high. 
However, \cite{Woosley04} showed, that these ignition conditions are only fulfilled for the first burst after the start of accretion on a pure ${}^{56}\textrm{Fe}$ atmosphere. Compositional inertia effects \citep{Taam93} for subsequent bursts significantly reduce peak temperature and the amount of hydrogen at ignition as ignition occurs at a lower pressure and depth thus limiting the rp-process to $A\lesssim 64$ with only a fraction of heavier isotopes produced.

The nuclear reaction network used in this work employs 304 isotopes (see table~[\ref{fig:sunet}]). All the connecting particle reactions are taken from the REACLIB (see \cite{Sakharuk06}). These reaction rates have also been used in \cite{Weinberg06}.  
The network includes all isotopes between the valley of stability and the proton dripline up to ${}^{64}\textrm{Ge}$. Here isotopes with $\beta^+$-half lives $> 1$ day are considered ``stable'' on the timescale of the burst intervals, so their daughters are not included. The hot proton-proton chains of \cite{Wiescher89a} are also included.
Above the ${}^{64}\textrm{Ge}$ waiting point, only isotopes between the proton drip line and half lives less than 1 minute are included. This is because protons only capture on these high-Z isotopes during the burst's peak temperature which is only sustained for a few seconds. 
Weak rates up to $Z=32$ are taken from \cite{Fuller80,Fuller82a,Fuller82b} and \cite{Langanke01}. Since only a small fraction of material is processed above $Z=32$, it is a reasonable approximation to ignore neutrino losses from heavier isotopes \citep{Schatz99}.
These considerations significantly reduce the size of the networkm which decreases the simulation run-time.
\begin{table}[tbph]
\centering
\begin{tabular}{|r|l|r|l|r|l|}
\hline
Z & A & Z & A & Z & A\\
\hline
\hline
n & 1       & Ar& 31--38  & Kr& 69--74\\
H & 1--3    & K&  35--39  & Rb& 73--77\\  
He& 3,4    & Ca& 36--44  & Sr& 74--78\\  
Li& 7       & Sc& 39--45  &  Y& 77--82\\  
Be& 7,8    & Ti& 40--47  & Zr& 78--83\\  
B & 8,11    & V&  43--49  & Nb& 81--85\\ 
C & 9,11,12 & Cr& 44--52  & Mo& 82--86\\ 
N & 12--15  & Mn& 47--53  & Tc& 85--88\\   
O & 13--18  & Fe& 48--56  & Ru& 86--91\\
F & 17--19  & Co& 51--57  & Rh& 89--93\\ 
Ne& 18--21  & Ni& 52--62  & Pd& 90--94\\ 
Na& 20--23  & Cu& 54--63  & Ag& 94--98\\
Mg& 21--25  & Zn& 55--66  & Cd& 95--99\\
Al& 22--27  & Ga& 59--67  & In& 98--104\\
Si& 24--30  & Ge& 60--68  & Sn& 99--105\\
P&  26--31  & As& 64--69  & Sb& 106 \\ 
S&  27--34  & Se& 65--72  & Te& 107 \\  
Cl& 30--35  & Br& 68--73  & & \\
\hline
\end{tabular}
\caption{The table shows the list of isotopes which describes the $rp$-process. See the main text for details. An earlier version of this reaction network has been used in the following works \citep{Fisker04,Fisker04b,Fisker05b,Fisker05a,Fisker06} but it now includes the hot proton-proton chains of \cite{Wiescher89a}. The network is described in more detail in \cite{Fisker06}.}\label{fig:sunet}
\end{table}
The inner boundary, i.e.~towards the neutron star crust, has been slightly improved compared to previous work which used either a massive substrate (\cite{Woosley04}) or parameter values (\cite{Rembges99}). This work uses the neutron star core code of \cite{Brown00,Brown03} which calculates the thermal luminosity emanating from the crust given the temperature at the atmosphere-ocean interface.
The code includes pair, photo, and plasmon neutrino emission. This neutrino luminosity is only a few percent of the thermal luminosity but still several orders of magnitude larger than the hydrodynamical uncertainty due to the conservative formulation of mechanical equations. 
Different types of convection occur when thermal fluctuations cause instabilities to grow. Their rate of growth determine the eddy-velocity, so all instabilities can be treated by the mixing length theory (MLT) implementation. The present work only includes the Schwarzschild-Ledoux instability, because it is the dominant form of convection during the burst (which is the only period relevant to this paper), whereas secular instabilities (e.g.~semi-convection) occur in between bursts and are negligible at high accretion rates, because the diffusion speed is smaller than the advection speed of the accretion. 
The initial model was computed by running the simulation for hundreds of bursts until the burst ashes had advected completely to the bottom of the model and the computational envelope was in thermal balance with the neutron star core model. At this point, the envelope was considered to be self-consistent, that is, independent of any possibly unphysical initial values, and a typical X-ray burst was picked for analysis. 

\section{Burst simulations}
\begin{figure}[tbph]
 \plotone{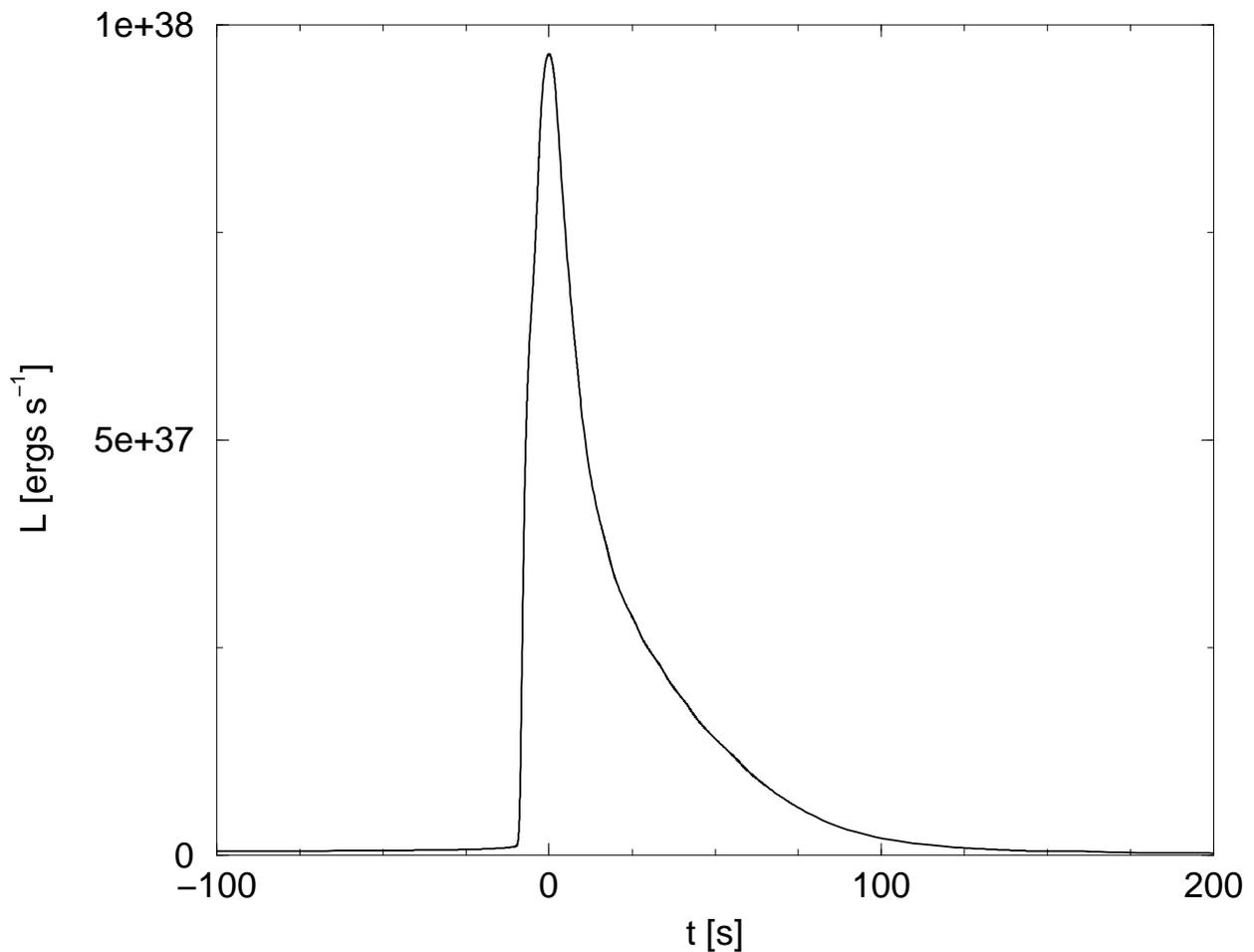}  
 \caption{The figure shows the luminosity as seen from an observer at infinity as a function of time for a typical burst. The timescale has been reset so that $t=0$ corresponds to the peak luminosity of the burst.}\label{fig:Lt} 
\end{figure}

Fig.~\ref{fig:Lt} shows the luminosity for the analyzed burst as a function of time. When comparing it to observations, it should be kept in mind that our model assumes a spherical symmetric ignition whereas in reality ignition most likely happens at a single point on the neutron star after which the flamefront spreads and eventually covers the entire neutron star. Therefore Fig.~\ref{fig:Lt} may be thought of as the luminosity of a single point under the assumption of negligible lateral heat transport. However, this assumption seems to provide a good comparisson to reality \citep{Galloway04}.

\begin{figure}[tbph]
 \plotone{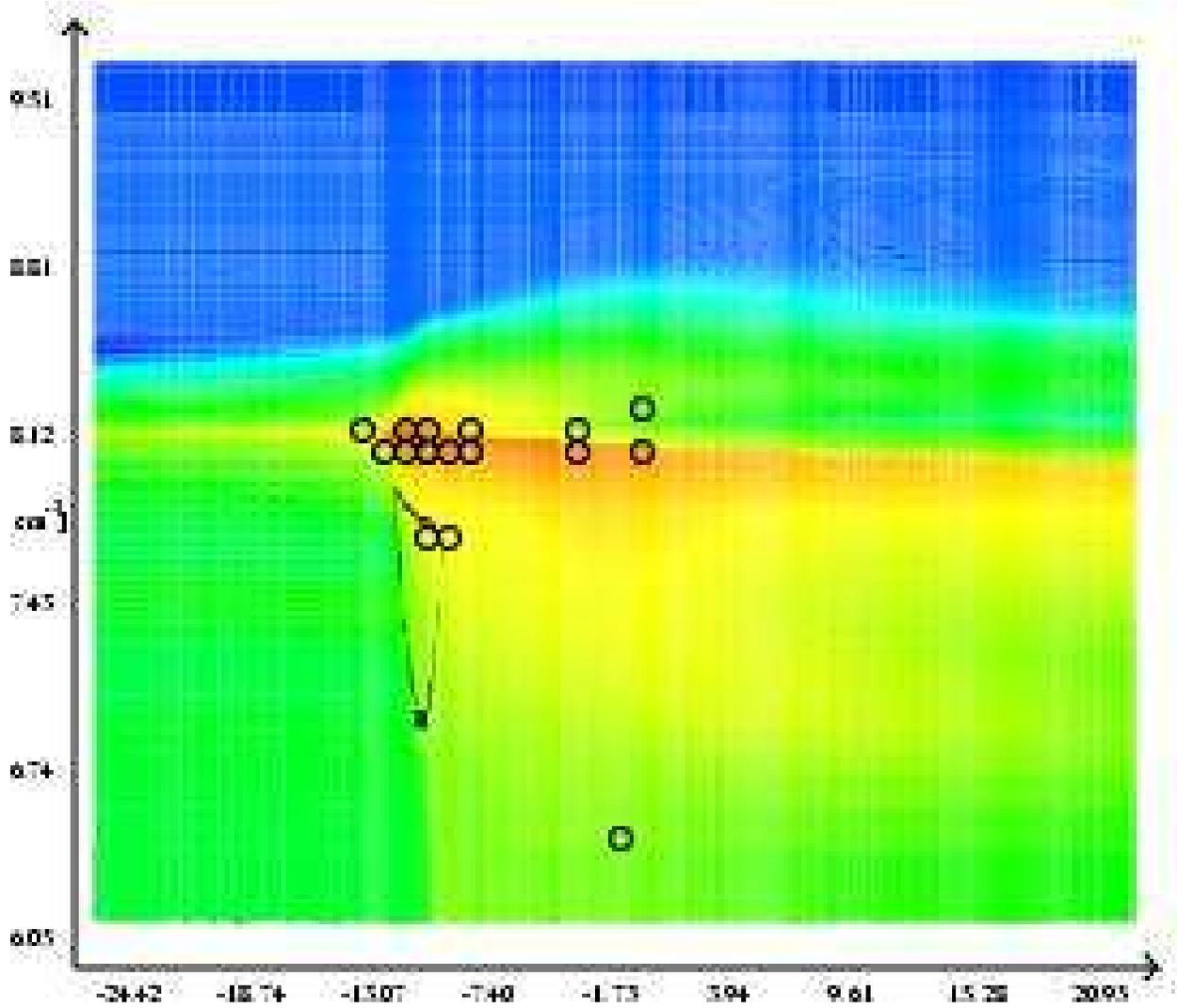} 
 \caption{The figure shows the specific nuclear energy release rate (color coded) as a function of logarithmic ($\log_{10}$) relativistic column density (see footnote~\ref{footnote:relativistic-y}) and time as well as the extent of the convective zone (black sail shaped outline). The black circles correspond to the descriptions of the reaction flow in \S\ref{sec:flow}. Starting from the top of the figure, they are: ocean (\ref{subsec:ocean}), ignition region (\ref{subsubsec:87chart1037}--\ref{subsubsec:87chart1375}, \ref{subsubsec:87chart971} is not shown), above ignition region (\ref{subsubsec:90chart1038}--\ref{subsubsec:90chart1483}), \ref{subsubsec:90chart971} and \ref{subsubsec:90chart1875} are not shown), bottom of the convective region (\ref{subsubsec:104chart1115}--\ref{subsubsec:104chart1214}), and surface (\ref{subsubsec:128chart1419}). 
}\label{fig:kippenhahn} 
\end{figure}

The luminosity is the combined product of the energy released from nuclear reactions at different depths. The Kippenhahn diagram in Fig.~\ref{fig:kippenhahn}  shows the specific nuclear energy release rate as a function of column density and time as well as the extent of the convective zone. 

The ignition region is easily indentified as a sudden and rapid increase increase in nuclear energy generation. This causes a temperature spike that causes a convective instability. Since convection transports heat very efficiently, the zones above the ignition point also ignite. This is clearly seen in Fig.~\ref{fig:kippenhahn}. However, this heat transport quickly restores the shallower radiative/conductive temperature gradient and convection quickly ends. The Kippenhahn diagram also demonstrates how nuclear energy generation decreases as fuel depletes and how residual helium from the previous burst contributes to nuclear reactions below the ignition region. Still, most of the nuclear energy is released, not at the point of ignition but in the hydrogen rich layers immediately above the ignition region. 

\section{Reaction flow}\label{sec:flow}
Runaways occurring in a mixed H/He layer mainly proceed via the $rp$-process (\cite{Wallace81}), where the characteristic timescale, $\tau_{rp}\sim\sum T_{1/2}$, is given by the sum of the half-lives of the $\beta$-decays in the reaction flow (\cite{Wormer94}). 
However, depending on the flow pattern, a simultaneuosly occurring $(\alpha,p)$-process, which does not depend on $\beta$-decays, may decrease the timescale through the $sd$-shell nuclei (\cite{Wallace81,Schatz98,Fisker04b}). 
As the runaway lasts several seconds for a H/He-ignited XRB, the temperature gradient only produces a minor convective instability.

\begin{figure}[tbph]
 \plotone{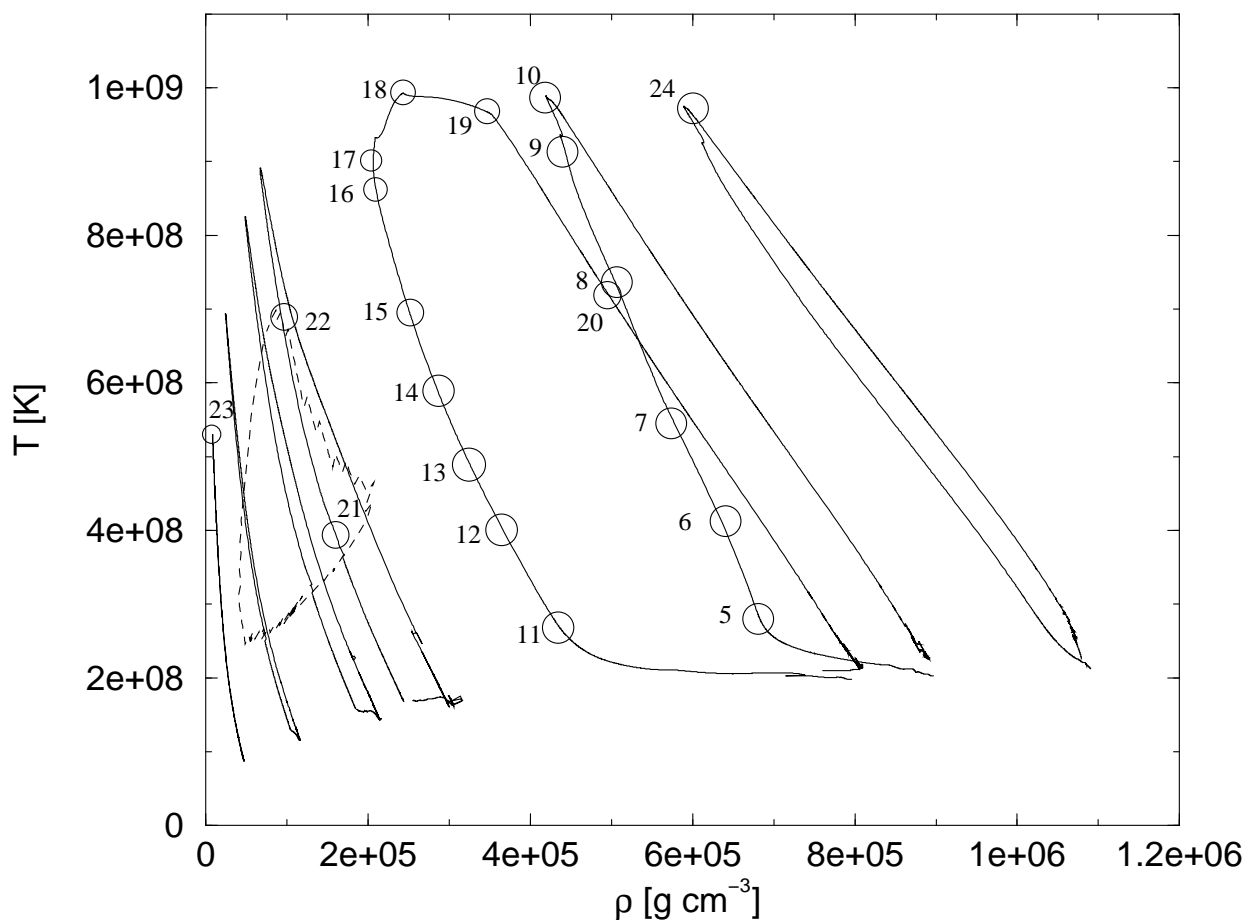} 
 \caption{From left to right (solid line): $y=1.8\times 10^6 \textrm{g}/\textrm{cm}^2$ (surface), $y=1.0\times 10^7 \textrm{g}/\textrm{cm}^2$ (top of the convective region), $y=2.5\times 10^7 \textrm{g}/\textrm{cm}^2$, $y=3.7\times 10^7 \textrm{g}/\textrm{cm}^2$ (bottom of the convective region), $y=7.9\times 10^7 \textrm{g}/\textrm{cm}^2$ (above ignition), $y=9.5\times 10^7 \textrm{g}/\textrm{cm}^2$ (ignition point), and $y=1.2\times 10^8 \textrm{g}/\textrm{cm}^2$ (ocean). The dashed line indicate the region which is convective during the rising of the burst. The circles and their associated numbers correspond to the figures in \S\ref{sec:flow}.}\label{fig:rhot} 
\end{figure}

The analysis of a one-dimensional X-ray burst model is very complex, as the model is characterized by rapidly changing temperature conditions and nuclear reaction sequences in each layer. All these effects are tightly interconnected through energy generation and heat transport by radiation and convection.
Since the different layers interact and also burn differently due to different compositions and temperature, the burst can not be understood based on the burning of one layer only, but must be analyzed for several different burning layers.
Therefore the analysis is split into four regions: the region around the ignition point, the convective region, the surface, and the ocean, which are sufficiently different to merit separate attention.
These regions are shown in Fig.~\ref{fig:rhot} which shows a trace of the burst conditions for different depths (pressures) during a complete revolution of the limit cycle. 

Following the cooling of the previous burst, the individual layers reach their lowest temperature and highest density of the cycle. The subsequent accretion increases the hydrogen content of the layer, which in turn lowers the density, because the increased electron abundance of hydrogen requires less mass to maintain the hydrostatic pressure (\cite{Joss77,Joss80}) compared to the heavier and more neutron-rich ashes (\cite{Hanawa84}). 
This is most clearly seen in Fig.~\ref{fig:rhot} just above the ignition region, which decreases its density by about a factor two during the quiescent phase, as the electron-rich surface ashes of the previous burst sink into this region.
For an accretion rate of $\dot{M}=1\cdot 10^{17}\textrm{g}/\textrm{s}$ and a recurrence time of $\Delta t \approx 11000\textrm{s}$, the neutron star accretes a mass of $\Delta M=\dot{M}\cdot\Delta t \approx 1.1\cdot10^{21}\textrm{g}$ $(5.5\cdot10^{-13}M_\odot)$ in between bursts. 
This means that matter above $y=5.8\times 10^7 \textrm{g}/\textrm{cm}^2$ is freshly accreted, whereas matter below comprise the old surface ashes of the previous burst(s), therefore the composition in the ignition region actually consist of heavier ashes with a comparably lower hydrogen/helium abundance. 
When the matter ignites and the nuclear runaway causes a rising temperature, it  eventually affects the degeneracy of the electrons and decreases the density further so the trace runs up the left leg of the cycle in Fig.~\ref{fig:rhot} until the fuel is exhausted as it burns into heavier ashes shortly after the peak temperature is reached. 
The $\beta^+$-decays during and subsequent to the $rp$-process decrease the electron abundance and bring the trace down the right leg as the envelope cools.
Therefore the separation in density between the rising leg and the decaying leg accounts for the change in composition, so the largest change happens around the ignition regions, whereas the surface does not change its composition much.
The different compositions and hydrostatic pressures with corresponding temperatures and densities of the burning regions change dynamically on a nuclear timescale defined as $\min(dt/d\ln Y_i)$, where $Y_i$ is the abundance of the $i$th isotope. 
Therefore the analysis of the nuclear reaction flow proceeds in a different way compared to previous works, which assumed solar abundances burning at fixed densities and temperatures and described the integrated flow over many minutes (\cite{Wormer94,Rembges97}); instead the instant flow rate is described as the thermodynamic state variables change.

The net reaction flow rate from isotope $i$ to isotope $j$ is defined by
\begin{equation}\label{eq:netreactionflow}
f_{ij}= -f_{ji}= \dot{Y}_{i\to j} - \dot{Y}_{j\to i}\,,
\end{equation}
where $\dot{Y}_{i\to j}$ is the time rate-of-change of the abundance of the $i$th isotope resulting from all reactions converting isotope $i$ to isotope $j$.
The flow-rates for the different times of Fig.~\ref{fig:rhot} will be demonstrated in the flowcharts of the following sections, which describe the reaction flow rates at the ignition point, the region above it, the convective zone, the surface (of our model), and the ashes going into the ocean.

In these figures the main reaction-flow is described by the heavy lines.
Very thick lines indicate $(p,\gamma)(\gamma,p)$-equilibrium.
Thin lines indicate a flow rate just above $10^{-6}\textrm{mol}/\textrm{g}/\textrm{s}$ increasing their thickness logarithmically to a maximum after which they stay constant. Also shown are the mass fractions, $X_A=\sum_{A_i=A} X_i$, for a given mass, $A$, as a function of mass number. If $X_A>0.20$ for any $A$, the bar is cut off and replaced with a dotted line. This only happens for the $A=1$ and $A=4$ cases where the concentration can be read in the figure caption.

\subsection{Ignition region}\label{subsec:ignition}
Between bursts, the surface ashes from the previous burst sink down under the weight of the newly accreting matter while the hot CNO cycle transforms hydrogen into helium. The hot CNO cycle is beta-limited and therefore burns at a fixed rate that mainly depends on the concentration of ${}^{14}\textrm{O}$ and ${}^{15}\textrm{O}$. It is partially moderated by a quiescent breakout via the ${}^{15}\textrm{O}(\alpha,\gamma){}^{19}\textrm{Ne}$ reaction which depletes ${}^{15}\textrm{O}$ and thus slows down the conversion of hydrogen into helium. This is discussed in more detail in \cite{Fisker06}. As seen in figure~\ref{fig:87chart971}, the reaction flow can subsequently return to the hot CNO cycle via the bi-cycle ${}^{19}\textrm{Ne}(\beta^+\nu)$ $(T_{1/2}=17.22\textrm{s})$ ${}^{19}\textrm{F}(p,\alpha)$ ${}^{16}\textrm{O}(p,\gamma)$  ${}^{17}\textrm{F}(p,\gamma)$ ${}^{18}\textrm{Ne}(\beta^+,\nu)$ $(T_{1/2}=1.672\textrm{s})$ ${}^{18}\textrm{F}(p,\alpha){}^{15}\textrm{O}$ which speeds up the conversion of hydrogen to helium and thus influences the composition for the runaway. This cycle is discussed in more detail in \cite{Cooper06}.

The hot CNO cycle increases the ${}^{4}\textrm{He}$ concentration until a runaway of the extremely temperature sensitive triple-alpha reaction ensues and causes a spike in the nuclear energy release rate.

\begin{figure}[tbph]
 \plotone{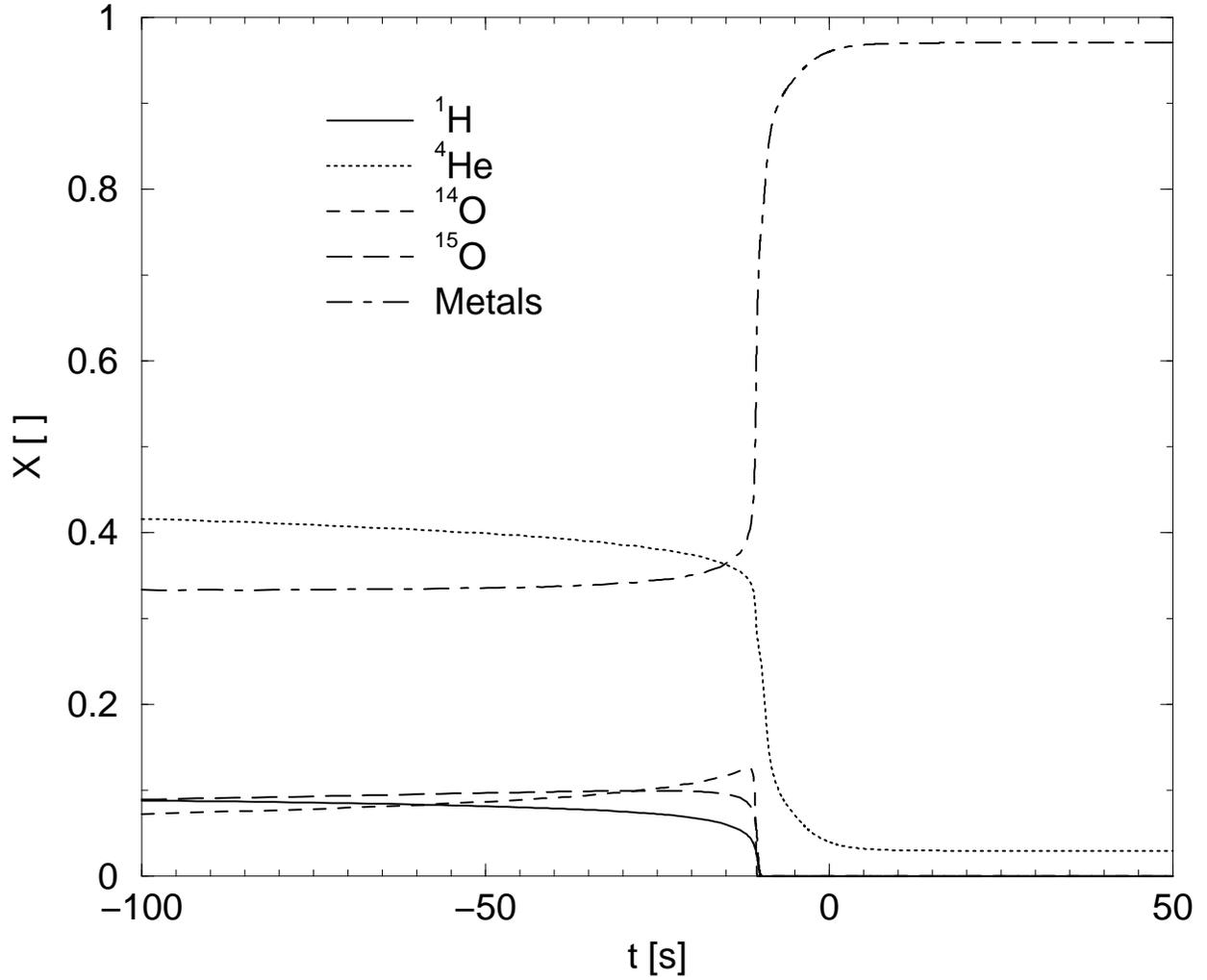} 
 \caption{The hydrogen, helium, CNO type matter, and metal (the rest) mass fractions as a function of time. The time scale has been synchronized to coincide with the burst luminosity peak at $t=0$. Notice the run-up in ${}^{14}\textrm{O}$ immediately prior to the runaway. Also note that the hydrogen in the ignition region is completely exhausted during the burst, while about $\sim 5\%$ helium remains.}\label{fig:xt_ignition} 
\end{figure}

The triple-alpha reaction creates ${}^{12}$C which immediately captures two protons to become ${}^{14}\textrm{O}$, causing the fraction of ${}^{14}\textrm{O}$ ($T_{1/2}=76.4$s) to increase as seen in Fig.~\ref{fig:xt_ignition}, since ${}^{14}\textrm{O}$ ($T_{1/2}=76.4$s) decays too slowly. 
Meanwhile the increasing temperature of the nascent nuclear runaway leads to a breakout of the hot CNO cycle into the $rp$-process. The details are described in the following subsections.

We now describe the reaction flow in terms of temperature, density, and proton and alpha fractions as the time develops. The time is synchronized, so $t=0$ coincides with the peak surface luminosity. 

\begin{figure}[tbph]
\centering
 \plotone{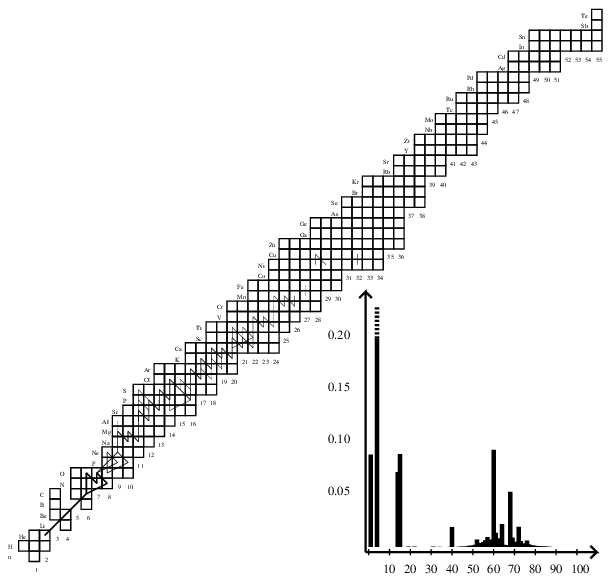} 
 \caption{Ignition: $T=2.86\cdot10^8\textrm{K}$, $\rho=6.81\cdot10^5\textrm{g}/\textrm{cm}^3$, $X=0.09$, $Y=0.42$, $t=-103.078\textrm{s}$. (see end of \S\ref{sec:flow} for an explanation of the diagram).}\label{fig:87chart971} 
\end{figure}

\subsubsection{Fig.~\ref{fig:87chart971}: $T=2.86\cdot10^8\textrm{K}$, $\rho=6.81\cdot10^5\textrm{g}/\textrm{cm}^3$, $X=0.09$, $Y=0.42$, $t=-103.078\textrm{s}$}\label{subsubsec:87chart971} 
At this time the increasing temperature has caused the flow rate of ${}^{15}\textrm{O}(\alpha,\gamma){}^{19}\textrm{Ne}$ (see \cite{Fisker06} for a detailed discussion of this rate) to reach $10\%$ of the ${}^{15}\textrm{O}(\beta^+,\nu){}^{15}\textrm{N}$ rate establishing a breakout of the hot CNO cycle (the $1\%$ limit was breached at $t=-558\textrm{s}$) which extends into the light iron region.

At this point proton captures establish a flow out of ${}^{19}\textrm{Ne}$. The matter in this flow can no longer return to the hot CNO cycle and the reaction flow proceeds with ${}^{19}\textrm{Ne}(p,\gamma)$ ${}^{20}\textrm{Na}(p,\gamma)$ ${}^{21}\textrm{Mg}$, where it is blocked by photodisintegration, because of the ${}^{21}\textrm{Mg}(p,\gamma)(\gamma,p){}^{22}\textrm{Al}$-equilibrium.  
Therefore the flow proceeds via ${}^{21}\textrm{Mg}(\beta^+,\nu)$ $(T_{1/2}=0.124\textrm{s})$ ${}^{21}\textrm{Na}(p,\gamma)$ ${}^{22}\textrm{Mg}(\beta^+,\nu)$ $(T_{1/2}=3.32\textrm{s})$ ${}^{22}\textrm{Na}(p,\gamma)$ ${}^{23}\textrm{Mg}(p,\gamma)$ ${}^{24}\textrm{Al}(p,\gamma)$ ${}^{25}\textrm{Si}$. 

Here the flow branches into either ${}^{25}\textrm{Si}(\beta^+,\nu)$ $(T_{1/2}=0.198\textrm{s})$ ${}^{25}\textrm{Al}(p,\gamma)$ ${}^{26}\textrm{Si}$ $(T_{1/2}=1.84\textrm{s})$ $(p,\gamma)$ ${}^{27}\textrm{P}$ or 
${}^{25}\textrm{Si}(p,\gamma)$ ${}^{26}\textrm{P}(\beta^+,\nu)$ $(T_{1/2}=0.020\textrm{s})$ ${}^{26}\textrm{Si}(p,\gamma)$ ${}^{27}\textrm{P}$ or 
${}^{25}\textrm{Si}(p,\gamma)$ ${}^{26}\textrm{P}(p,\gamma)$ ${}^{27}\textrm{S}(\beta^+,\nu)$ ${}^{27}\textrm{P}$ all of which have ${}^{27}\textrm{P}$ as the end point. The characteristic time depends on the mass fraction weighed harmonic mean of the half lives of the beta decays along respective pathways. As the temperature rises, the proton capture branches become initially more dominant but then decrease again as photodisintegration of the weakly proton bound and short lived proton-rich $\textrm{P}$ and $\textrm{S}$ isotopes steers the flow away from the dripline again. Yet at this particular temperature the beta decay path of ${}^{25}\textrm{Si}$ dominates the proton capture to ${}^{26}\textrm{P}$. 

From this point, ${}^{27}\textrm{P}(\beta^+,\nu)$ $(T_{1/2}=0.242\textrm{s})$ ${}^{27}\textrm{Si}(p,\gamma)$ ${}^{28}\textrm{P}(p,\gamma)$ ${}^{29}\textrm{S}(\beta^+,\nu)$ $(T_{1/2}=0.146\textrm{s})$ ${}^{29}\textrm{P}(p,\gamma)$ ${}^{30}\textrm{S}$ which has a half life of $1.07\textrm{s}$. The $Q$-value of proton capture on ${}^{30}\textrm{S}$ is only $290.6\textrm{keV}$ which makes ${}^{31}\textrm{Cl}$ subject to photodisintegration at higher temperatures. At later times, this can have a large effect on the observed luminosity of the burst \citep{Fisker04b}. 

Here, the proton capture still dominates, so the flow proceeds via ${}^{30}\textrm{S}(p,\gamma)$ $(T_{1/2}=1.07\textrm{s})$ ${}^{31}\textrm{Cl}(\beta^+,\nu)$ $(T_{1/2}=0.270\textrm{s})$ ${}^{31}\textrm{S}$. This $(T_{1/2}=2.13 \textrm{s})$ isotope either beta decays and returns to ${}^{28}\textrm{Si}$ via ${}^{31}\textrm{P}(p,\alpha)$ ${}^{28}\textrm{Si}$ or captures a proton and proceeds via  
${}^{31}\textrm{S}(p,\gamma)$ (\cite{Iliadis99}) 
${}^{32}\textrm{Cl}(\beta^+,\nu)$ $(T_{1/2}=0.285\textrm{s})$ 
${}^{32}\textrm{S}(p,\gamma)$ 
${}^{33}\textrm{Cl}(p,\gamma)$ 
${}^{34}\textrm{Ar}(\beta^+,\nu)$ $(T_{1/2}=0.811\textrm{s})$
${}^{34}\textrm{Cl}(p,\gamma)$ 
${}^{35}\textrm{Ar}(p,\gamma)$ (\cite{Iliadis99}) 
${}^{36}\textrm{K}(\beta^+,\nu)$ $(T_{1/2}=0.302\textrm{s})$
${}^{36}\textrm{Ar}(p,\gamma)$ 
${}^{37}\textrm{K}(p,\gamma)$ 
${}^{38}\textrm{Ca}$. 

Since ${}^{39}\textrm{Sc}$ and ${}^{40}\textrm{Sc}$ are almost proton unbound the flow must wait for ${}^{38}\textrm{Ca}$ $(T_{1/2}=0.416\textrm{s})$ and ${}^{39}\textrm{Ca}$ $(T_{1/2}=0.799\textrm{s})$ to $\beta^+$-decay before the flow stops at the well-bound ${}^{40}\textrm{Ca}$ isotope.
A CaScTi cycle exists on the well-bound ${}^{40}\textrm{Ca}$ so that ${}^{40}\textrm{Ca}(p,\gamma)$ ${}{}^{41}\textrm{Sc}(p,\gamma)$ ${}^{42}\textrm{Ti}(\beta^+,\nu)$ $(T_{1/2}=0.189\textrm{s})$ ${}^{42}\textrm{Sc}(p,\gamma)$ ${}^{43}\textrm{Ti}(\beta^+,\nu)$ $(T_{1/2}=0.429\textrm{s})$ ${}^{43}\textrm{Sc}(p,\alpha)$ ${}^{40}\textrm{Ca}$.
The breakout from this cycle happens from ${}^{43}\textrm{Sc}$ which proceeds to capture protons going through ${}^{44}\textrm{Ti}$ and ${}^{45}\textrm{V}$ until it ends at ${}^{52}\textrm{Fe}$.

The total timescale for this sequence is (c.f.~\cite{Wormer94}) $\tau=\ln(2)^{-1}\sum T_{1/2} \sim 8\textrm{s}$, which is slower than the time it takes to cover the star with a deflagration wave by a factor four \citep{Fryxell82b}. Therefore a one-dimensional approximation is still reasonable. 
Later when the $(\alpha,p)$-process ignites and the temperature increases, the reaction flow will move closer to the dripline decreasing the $\beta$-half-lives, thus making the timescales comparable. At that point, our model is no longer fully predictive of hydrodynamically influenced (extensive) observables such as the time-dependent luminosity. However, our model still provides a local (intensive) description of the burning conditions, and therefore a realistic description of the reaction flow.

\begin{figure}[tbph]
\centering
 \plotone{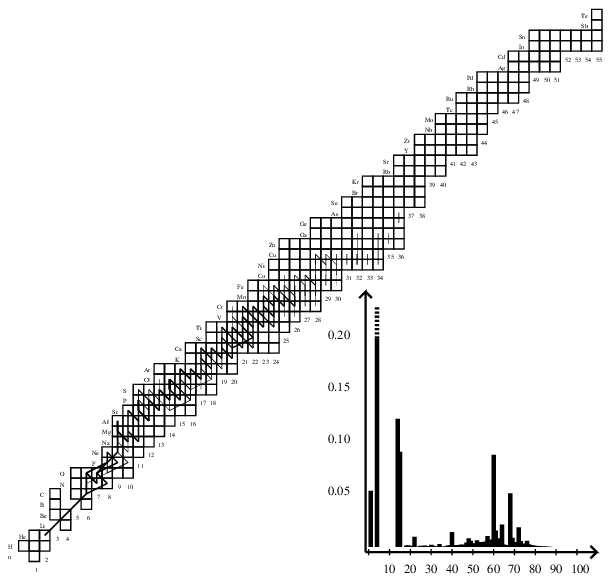} 
 \caption{Ignition: $T=3.99\cdot10^8\textrm{K}$, $\rho=6.41\cdot10^5\textrm{g}/\textrm{cm}^3$, $X=0.05$, $Y=0.36$, $t=-12.938\textrm{s}$. (see end of \S\ref{sec:flow} for an explanation of the diagram).}\label{fig:87chart1037} 
\end{figure}

\subsubsection{Fig.~\ref{fig:87chart1037}: $T=3.99\cdot10^8\textrm{K}$, $\rho=6.41\cdot10^5\textrm{g}/\textrm{cm}^3$, $X=0.05$, $Y=0.36$, $t=-12.938\textrm{s}$}\label{subsubsec:87chart1037}
Approximately 90 seconds later the ${}^{14}\textrm{O}(\alpha,p){}^{17}\textrm{F}$ reaction reaches $1/3$ of the flow rate of the ${}^{15}\textrm{O}(\alpha,\gamma){}^{19}\textrm{Ne}$-reaction. This starts a second hot CNO bi-cycle: ${}^{14}\textrm{O}(\alpha,p)$ ${}^{17}\textrm{F}(p,\gamma)$ ${}^{18}\textrm{Ne}(\beta^+,\nu)$ ${}^{18}\textrm{F}(p,\alpha){}^{15}\textrm{O}$ which runs alongside the bi-cycle discussed above in \S\ref{subsec:ignition}.

 At this stage ${}^{22}\textrm{Mg}(p,\gamma){}^{23}\textrm{Al}$  and ${}^{22}\textrm{Mg}(\beta^+,\nu)$ $( T_{1/2}=3.34\textrm{s}){}^{22}\textrm{Na}$ become comparable. 
Consequently the flow path through ${}^{22}\textrm{Mg}(p,\gamma)$ ${}^{23}\textrm{Al}(p,\gamma)$ ${}^{24}\textrm{Si}(\beta^+,\nu)$ $( T_{1/2}=0.190\textrm{s})$ ${}^{24}\textrm{Al}$ competes with ${}^{22}\textrm{Mg}(\beta^+,\nu)$ $( T_{1/2}=3.34\textrm{s})$ ${}^{22}\textrm{Na}(p,\gamma)$ ${}^{23}\textrm{Mg}(p,\gamma)$ ${}^{24}\textrm{Al}$ effectively creating a shortcut. Since the flow rates are about equal, the effective timescale becomes the flow rate weighted harmonic mean of the two half-lives $\approx 0.10\textrm{s}$, which is much faster than before. This reduces the total timescale to reach ${}^{40}\textrm{Ca}$ to $\sim 5\textrm{s}$.  
A similar shortcut exists with  ${}^{25}\textrm{Si}(p,\gamma)$ ${}^{26}\textrm{P}(p,\gamma)$ ${}^{27}\textrm{S}(\beta^+,\nu)$ $( T_{1/2}=0.021\textrm{s})$ ${}^{27}\textrm{P}$ competing with ${}^{25}\textrm{Si}(\beta^+,\nu)$ $( T_{1/2}=0.188\textrm{s})$ ${}^{25}\textrm{Al}(p,\gamma)$ ${}^{26}\textrm{Si}(p,\gamma)$ ${}^{27}\textrm{P}$ however, here the proton capture $Q$-value is only $141\textrm{keV}$, so the faster path is reduced by photodisintegration. 

At this time, the concentration of ${}^{31}\textrm{Cl}$ has peaked and is now being destroyed by photodisintegration. Therefore the flow must pass through the ${}^{30}\textrm{S}(\beta^+,\nu)$ $( T_{1/2}=1.08\textrm{s})$ reaction, which is the slowest weak reaction in the flow and adds about a second to the total timescale. 

Reaching ${}^{31}\textrm{S}$ the flow now branches again. Instead of going through the slower ${}^{31}\textrm{S}(\beta^+,\nu)$ $(T_{1/2}=2.13 \textrm{s})$ ${}^{31}\textrm{P}(p,\gamma)$  ${}^{32}\textrm{S}(p,\gamma)$  ${}^{33}\textrm{Cl}$, the flow can now go directly through either ${}^{31}\textrm{S}(p,\gamma)$ ${}^{32}\textrm{Cl}(\beta^+,\nu)$ $(T_{1/2}=0.293 \textrm{s})$ ${}^{32}\textrm{S}(p,\gamma)$  ${}^{33}\textrm{Cl}$ or ${}^{31}\textrm{S}(p,\gamma)$ ${}^{32}\textrm{Cl}(p,\gamma)$ ${}^{33}\textrm{Ar}(\beta^+,\nu)$ $(T_{1/2}=0.153 \textrm{s})$ ${}^{33}\textrm{Cl}$ which shaves another 2 seconds off the characteristic time for the $rp$-process.

The flow now breaks into the $pf$-shell nuclei by proton-captures on ${}^{39}\textrm{Ca}$ and ${}^{40}\textrm{Ca}$ (\cite{Wiescher89b}). 
The very fast $\beta^+$-decays on the highly radioactive Sc and Ti isotopes cause the flow to spread (Fig.\ref{fig:87chart1037}) and makes an analysis of the timescales difficult. 

The hot CNO like cycle 
discussed in the previous section now has proton capture breakouts on via ${}^{42}\textrm{Ti}(p,\gamma)$ ${}^{43}\textrm{V}$ and ${}^{43}\textrm{Ti}(p,\gamma)$ ${}^{44}\textrm{V}$ and ${}^{43}\textrm{Sc}(p,\gamma)$ ${}^{44}\textrm{Ti}$.

These reaction pass through ${}^{43}\textrm{V}$ and ${}^{44}\textrm{V}$ and after several combination of proton capture and beta decays, the flow goes through the ${}^{45}\textrm{V}$ bottleneck which can only happen through either a $\beta^+$-decay $(T_{1/2}=0.59\textrm{s})$ or a proton capture to ${}^{46}\textrm{Cr}$. 
The next bottleneck is ${}^{48}\textrm{Cr}$ which can be reached from ${}^{46}\textrm{V}$ by either ${}^{46}\textrm{V}(\beta^+,\nu)$ $(T_{1/2}=0.429\textrm{s})$ ${}^{46}\textrm{Ti}(p,\gamma)$ ${}^{47}\textrm{Ti}(p,\gamma)$ ${}^{48}\textrm{Cr}$ or ${}^{46}\textrm{V}(p,\gamma)$ ${}^{47}\textrm{Cr}(\beta^+,\nu)$ $(T_{1/2}=0.497\textrm{s})$ ${}^{47}\textrm{V}(p,\gamma)$ ${}^{48}\textrm{Cr}$ or ${}^{46}\textrm{V}(p,\gamma)$ ${}^{47}\textrm{Cr}(p,\gamma)$ ${}^{48}\textrm{Mn}(\beta^+,\nu)$ $(T_{1/2}=0.030\textrm{s})$ ${}^{48}\textrm{Cr}$. 

The ${}^{48}\textrm{Cr}$ bottleneck has a half life of $T_{1/2}=2.02\textrm{h}$ which makes it ``stable'' on the timescale of the burst. The ${}^{48}\textrm{Cr}(p,\gamma)$ ${}^{49}\textrm{Mn}$ reaction is therefore important at this stage because it is the only way for the flow to proceed. 

After capturing a proton the flow proceeds from ${}^{49}\textrm{Mn}$ to ${}^{50}\textrm{Mn}$ via either a beta decay followed by a proton capture or vice versa. The flow from ${}^{50}\textrm{Mn}$ to ${}^{51}\textrm{Mn}$ proceeds in a similar manner. The ${}^{51}\textrm{Mn}$ isotope has a half life of $T_{1/2}=35.3\textrm{m}$ so ${}^{51}\textrm{Mn}$ captures a proton and becomes ${}^{52}\textrm{Fe}$. 

There is a slight flow out of ${}^{52}\textrm{Fe}$ that moves to ${}^{56}\textrm{Fe}$ via a series of proton captures followed by beta decays. At this point ${}^{56}\textrm{Fe}$ captures several protons to ${}^{59}\textrm{Cu}$ which decays and proton captures and decays to ${}^{60}\textrm{Ni}$. There are also proton captures on Ni-Zn ashes from the previous burst at this time.

\begin{figure}[tbph]
\centering
\plotone{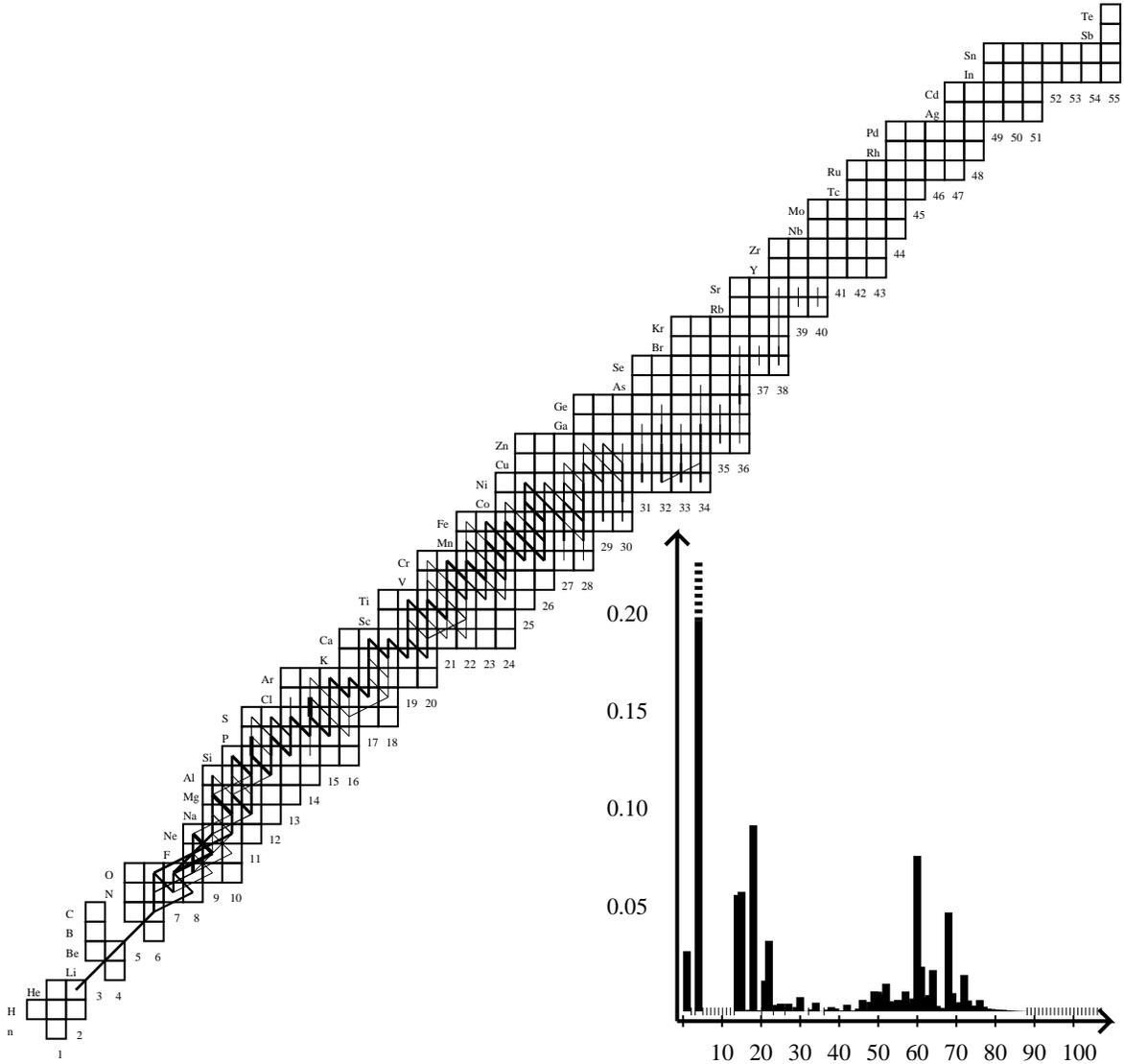} 
 \caption{Ignition: $T=5.37\cdot10^8\textrm{K}$, $\rho=5.81\cdot10^5\textrm{g}/\textrm{cm}^3$, $X=0.03$, $Y=0.31$, $t=-10.631\textrm{s}$. (see end of \S\ref{sec:flow} for an explanation of the diagram).}\label{fig:87chart1068} 
\end{figure}

\subsubsection{Fig.~\ref{fig:87chart1068}: $T=5.37\cdot10^8\textrm{K}$, $\rho=5.81\cdot10^5\textrm{g}/\textrm{cm}^3$, $X=0.03$, $Y=0.31$, $t=-10.631\textrm{s}$}\label{subsubsec:87chart1068} 
At this point the ${}^{14}\textrm{O}(\alpha,p)$ ${}^{17}\textrm{F}$ reaction is 5 times stronger than the ${}^{15}\textrm{O}(\alpha,\gamma){}^{19}\textrm{Ne}$ reaction. This starts the $(\alpha,p)$-proces (\cite{Wallace81,Schatz98,Fisker04b}) which here runs as ${}^{14}\textrm{O}(\alpha,p)$ ${}^{17}\textrm{F}(p,\gamma)$ ${}^{18}\textrm{Ne}(\alpha,p)$ ${}^{21}\textrm{Na}(p,\gamma)$ ${}^{22}\textrm{Mg}(\alpha,p)$ ${}^{25}\textrm{Al}(p,\gamma)$ ${}^{26}\textrm{Si}$. 

Another $(\alpha,p)$-reaction exists on ${}^{21}\textrm{Mg}(\alpha,p)$ ${}^{24}\textrm{Al}$. This reaction soon overpowers the ${}^{22}\textrm{Al}$ $\beta^+$-decay from the ${}^{21}\textrm{Mg}(p,\gamma)(\gamma,p)$ ${}^{22}\textrm{Al}$ equilibrium which becomes largely irrelevant for the burst from this point. 

While ${}^{31}\textrm{Cl}$ above ${}^{30}\textrm{S}$ is in $(p,\gamma)(\gamma,p)$-equilibrium with ${}^{32}\textrm{Ar}$, the main flow goes through the beta decay of ${}^{30}\textrm{S}$ which has a significant impact on the energy generation as it blocks the rest of the flow of the$rp$-process.

In the Ca-Fe region, the flow moves closer to the dripline with ${}^{43}\textrm{V}(p,\gamma)$ ${}^{44}\textrm{Cr}(\beta^+,\nu)$ $(T_{1/2}=0.030\textrm{s})$ ${}^{44}\textrm{V}$ and ${}^{47}\textrm{Mn}(p,\gamma)$ ${}^{48}\textrm{Fe}(\beta^+,\nu)$ $(T_{1/2}=0.030\textrm{s})$ ${}^{48}\textrm{Mn}$ becoming more active. 

Above Fe, the flow extends to ${}^{60}\textrm{Zn}$ as more proton rich nuclei start to capture protons. With ${}^{60}\textrm{Zn}$'s $T_{1/2}=4.10\textrm{m}$ halflife, ${}^{59}\textrm{Cu}$'s $T_{1/2}=83.1\textrm{s}$ halflife, and the stable ${}^{58}\textrm{Ni}$, the ${}^{60}\textrm{Zn}(p,\gamma)$ ${}^{61}\textrm{Ga}$ reaction is the only way to move the flow forward. 

In addition, proton captures on heavier isotopes from the previous burst moves towards the dripline. This compositional inertia increases the average mass and charge of the final ashes.

\begin{figure}[tbph]
\centering
\plotone{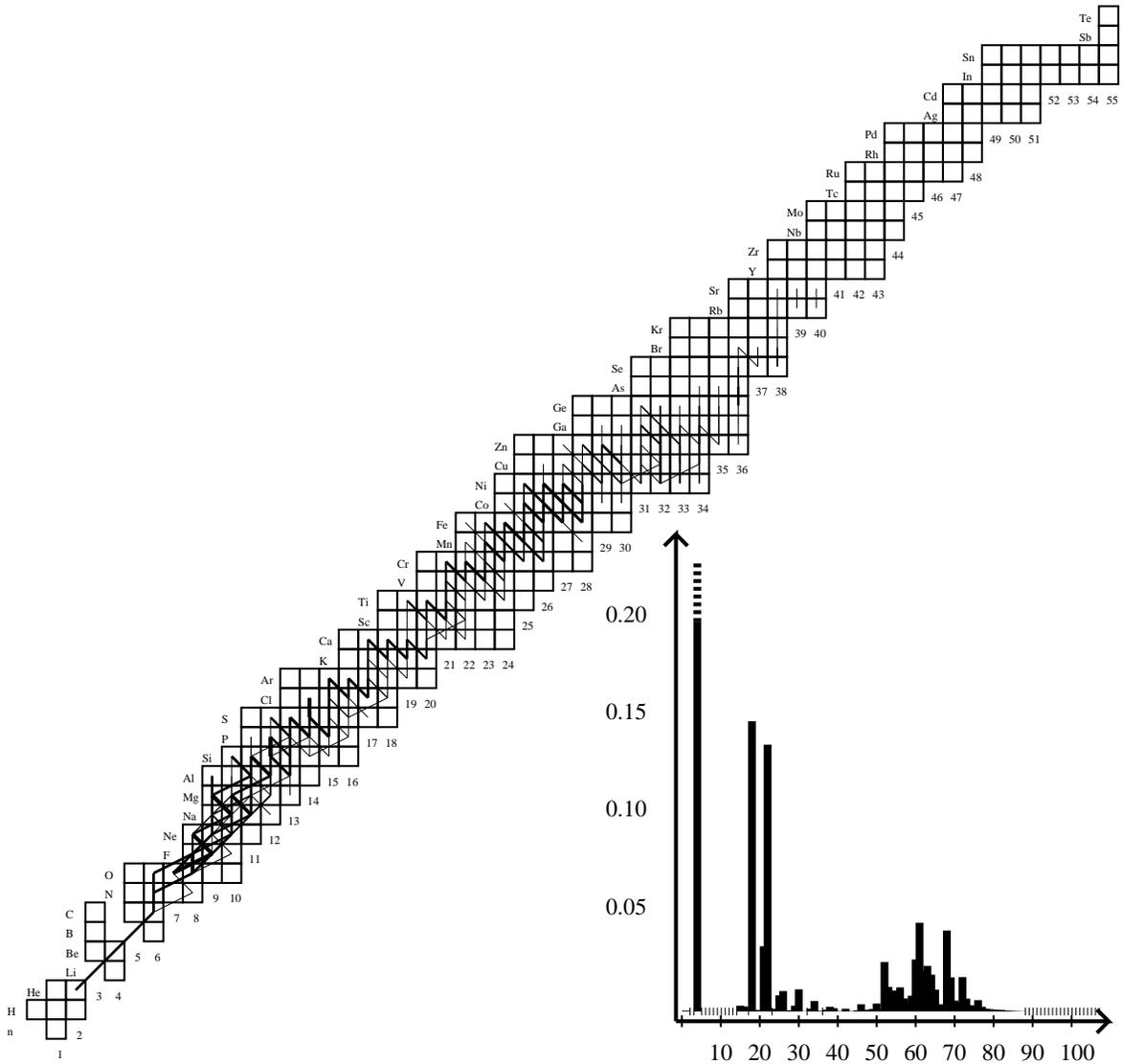} 
 \caption{Ignition: $T=7.30\cdot10^8\textrm{K}$, $\rho=5.07\cdot10^5\textrm{g}/\textrm{cm}^3$, $X=3.5\times 10^{-4}$, $Y=0.26$, $t=-9.980\textrm{s}$. (see end of \S\ref{sec:flow} for an explanation of the diagram).}\label{fig:87chart1145} 
\end{figure}

\subsubsection{Fig.~\ref{fig:87chart1145}: $T=7.30\cdot10^8\textrm{K}$, $\rho=5.07\cdot10^5\textrm{g}/\textrm{cm}^3$, $X=3.5\times 10^{-4}$, $Y=0.26$, $t=-9.980\textrm{s}$}\label{subsubsec:87chart1145}
The protons are now almost exhausted. This can also be seen in Fig.~\ref{fig:xt_ignition}. Once this happens, the proton-rich isotopes near the driplines decay towards the valey of stability, where they undergo $(\alpha,p)$-reactions. The figure shows the situation with less than $1\%$ hydrogen so proton captures are still occurring although they are weakening. Also, hydrogen from $(\alpha,p)$-reactions can still serve as a catalyst in the $(\alpha,p)$-process.

The $(\alpha,p)$-process extends further and now includes the ${}^{25}\textrm{Si}(\alpha,p)$ ${}^{28}\textrm{P}$ and ${}^{26}\textrm{Si}(\alpha,p)$ ${}^{29}\textrm{P}$ reactions. 

SiPS, PSCl, ArKCa, and CaScTi cycles are observed with ${}^{28}\textrm{Si}$, ${}^{32}\textrm{P}$, ${}^{36}\textrm{Ar}$ and ${}^{40}\textrm{Ca}$ as the nexus, but they are not consequential to the flow.

In the Zn region, the flow proceeds via ${}^{59}\textrm{Cu}(\beta^+,\nu)$ $(T_{1/2}=84.5\textrm{s})$ as the proton captures on ${}^{60}\textrm{Zn}$ remain weak. This flow reaches  ${}^{63}\textrm{Ge}$ and  ${}^{64}\textrm{Ge}$ but will not move further in this region as the hydrogen concentration is rapidly depleting.

\begin{figure}[tbph]
\centering
 \plotone{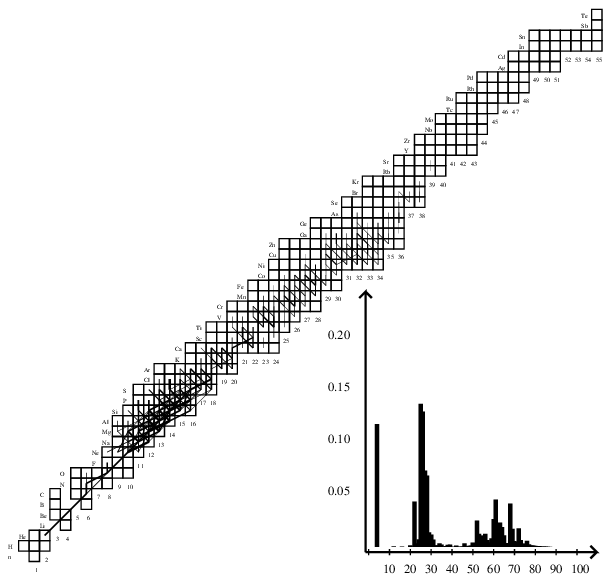} 
 \caption{Ignition: $T=9.03\cdot10^8\textrm{K}$, $\rho=4.46\cdot10^5\textrm{g}/\textrm{cm}^3$, $X3.3\times 10^{-6}$, $Y=0.12$, $t=-8.075\textrm{s}$. (see end of \S\ref{sec:flow} for an explanation of the diagram).}\label{fig:87chart1225} 
\end{figure}

\subsubsection{Fig.~\ref{fig:87chart1225}: $T=9.03\cdot10^8\textrm{K}$, $\rho=4.46\cdot10^5\textrm{g}/\textrm{cm}^3$, $X=3.3\times 10^{-6}$, $Y=0.12$, $t=-8.075\textrm{s}$}\label{subsubsec:87chart1225} 
At this point the region receives more heat from adjacent regions than it produces. This allows endothermic $(\alpha,p)$-reactions on better bound nuclei.

Presently, sufficient material has been moved to the ${}^{60}\textrm{Zn}$ isotope ensuring its decay and extending the $rp$-process from light isotopes into heavier isotopes. However, the only protons available for capture on these heavy nuclei come from $(\alpha,p)$-reactions on $A=20$--$36$ so the general lack of protons ensures that the $rp$-process at this point does not proceed beyond $A=64$. Additionally the shortage of protons means that the flow moves away from the proton dripline with the remaining protons generally capturing on the currently most abundant nuclei (now determined by half-life) with the largest cross sections and the lowest Coulomb barriers, that is, $A=20$--$36$.

\begin{figure}[tbph]
\centering
 \plotone{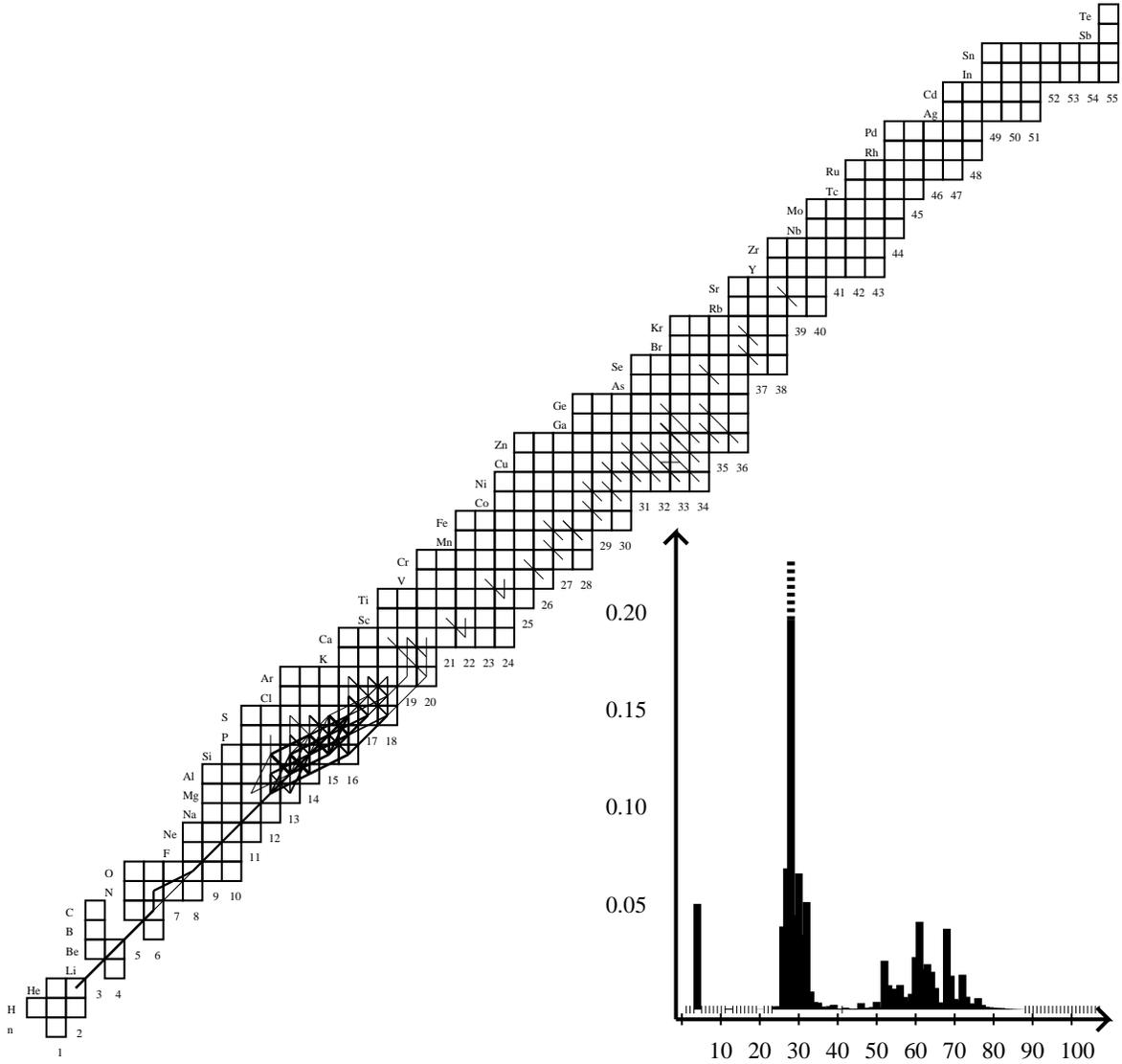} 
 \caption{Ignition: $T=9.89\cdot10^8\textrm{K}$, $\rho=4.20\cdot10^5\textrm{g}/\textrm{cm}^3$, $X=2.4\times 10^{-9}$, $Y=0.05$, $X_{28}=0.22$, $t=-3.018\textrm{s}$. (see end of \S\ref{sec:flow} for an explanation of the diagram).}\label{fig:87chart1375} 
\end{figure}

\subsubsection{Fig.~\ref{fig:87chart1375}: $T=9.89\cdot10^8\textrm{K}$, $\rho=4.20\cdot10^5\textrm{g}/\textrm{cm}^3$, $X=2.4\times 10^{-9}$, $Y=0.05$, $X_{28}=0.22$, $t=-3.018\textrm{s}$}\label{subsubsec:87chart1375} 
The flow through the alpha-chain nuclei is clearly seen in Fig.~\ref{fig:87chart1375} which shows the reaction flow at the time where maximum temperature is achieved. 
Notice that ${}^{12}\textrm{C}(p,\gamma)$ ${}^{13}\textrm{N}(\alpha,p)$ ${}^{16}\textrm{O}$ is much stronger than the direct ${}^{12}\textrm{C}(\alpha,\gamma)$ ${}^{16}\textrm{O}$-reaction as long as the $(\alpha,p)$-reactions are still possible on heavier isotopes \citep{Weinberg06}.

The reaction flow continues with $(\alpha,\gamma)$-reactions up to  ${}^{36}\textrm{Ar}$. Eventually the downward heat flux from the upper regions becomes too weak to sustain the $(\alpha,p)$-reactions and the reactions die out leaving only radioactive isotopes, which slowly decay to stabler ones.

\subsection{Above the ignition region}
It is relevant to know the reaction flow and its energy release at the depth, which reach the highest temperature during the burst, because it heats up adjacent and colder regions. This is because heat is transported as electrons and photons are diffused along a negative temperature gradient.
The highest temperature of a burst ignited by mixed hydrogen and helium is reached just above the point of point of ignition.

\begin{figure}[tbph]
\centering
 \plotone{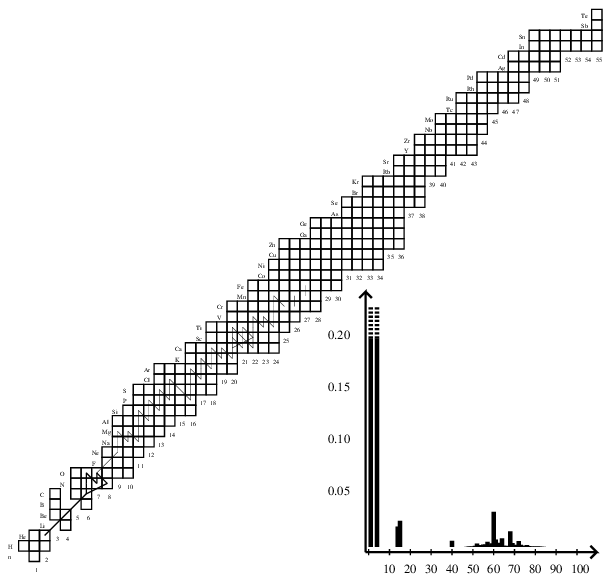} 
 \caption{Above ignition: $T=2.80\cdot10^8\textrm{K}$, $\rho=4.29\cdot10^5\textrm{g}/\textrm{cm}^3$, $X=0.41$, $Y=0.43$, $t=-103.078\textrm{s}$. (see end of \S\ref{sec:flow} for an explanation of the diagram).}\label{fig:90chart971} 
\end{figure}

\subsubsection{Fig.~\ref{fig:90chart971}: $T=2.80\cdot10^8\textrm{K}$, $\rho=4.29\cdot10^5\textrm{g}/\textrm{cm}^3$, $X=0.41$, $Y=0.43$, $t=-103.078\textrm{s}$}\label{subsubsec:90chart971}
The ${}^{15}\textrm{O}(\alpha,\gamma){}^{19}\textrm{Ne}$-reaction is less important at this depth, because less ``hot CNO material'' has been created due to lower operating temperatures and densities of the triple-alpha process. 
So while the reaction burns off the existing ${}^{15}\textrm{O}$, the runaway at this depth occurs, when the heat from the ignition point below increases the $3\alpha$ reaction, so the ${}^{12}\textrm{C}(p,\gamma)$ ${}^{13}\textrm{N}(p,\gamma)$ ${}^{14}\textrm{O}(\alpha,p)$ ${}^{17}\textrm{F}$ reaction path dominates the eventual runaway. This is because ${}^{14}\textrm{O}$ ($T_{1/2}=70.6$s) does not have the time to decay during the runaway. However, at this point the ${}^{15}\textrm{O}(\alpha,\gamma){}^{19}\textrm{Ne}$-reaction does establish a very weak flow to the iron-region along a flow path identical to the initial path in the deeper region. 

Although this region contains former surface ashes there are no further captures beyond these isotopes at the present time. 

\begin{figure}[tbph]
\centering
 \plotone{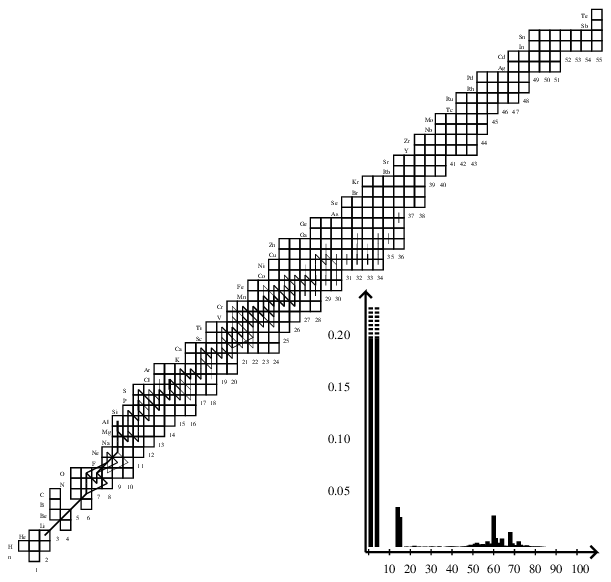} 
 \caption{Above ignition: $T=3.97\cdot10^8\textrm{K}$, $\rho=3.68\cdot10^5\textrm{g}/\textrm{cm}^3$, $X=0.400$, $Y=0.406$, $t=-12.387\textrm{s}$. (see end of \S\ref{sec:flow} for an explanation of the diagram).}\label{fig:90chart1038} 
\end{figure}

\subsubsection{Fig.~\ref{fig:90chart1038}: $T=3.97\cdot10^8\textrm{K}$, $\rho=3.68\cdot10^5\textrm{g}/\textrm{cm}^3$, $X=0.400$, $Y=0.406$, $t=-12.387\textrm{s}$}\label{subsubsec:90chart1038}

At this point ${}^{14}\textrm{O}(\alpha,p){}^{17}\textrm{F}$ reaction reaches $1/3$ of the flow rate of the ${}^{15}\textrm{O}(\alpha,\gamma){}^{19}\textrm{Ne}$-reaction. This happens at a lower temperature for this depth, because the ${}^{14}\textrm{O}$/${}^{15}\textrm{O}$ abundance-ratio is relatively higher.

The breakout establishes the second hot CNO bi-cycle (discussed in \S\ref{subsubsec:87chart1037}); in contrast to the ignition point, the first bi-cycle (discussed in \S\ref{subsec:ignition}) is not established, because it is already sufficiently hot and there is sufficient hydrogen to capture on ${}^{19}\textrm{Ne}$ destroying it immediately.

The heat flux building up rapidly from the ignition point below means that short cuts e.g.~${}^{22}\textrm{Mg}(p,\gamma)$ ${}^{23}\textrm{Al}(p,\gamma)$ ${}^{24}\textrm{Si}(\beta^+,\nu)$ $( T_{1/2}=0.190\textrm{s})$ ${}^{24}\textrm{Al}$ competing with ${}^{22}\textrm{Mg}(\beta^+,\nu)$ $( T_{1/2}=3.37\textrm{s})$ ${}^{22}\textrm{Na}(p,\gamma)$ ${}^{23}\textrm{Mg}(p,\gamma)$ ${}^{24}\textrm{Al}$ quickly becomes dominated by the leg closest to the proton dripline. 

From this point the reactions are identical to the flow described in \S\ref{subsubsec:87chart1037}. Since there is more hydrogen in this region the reaction on isotopes heaver than $\textrm{Mn}$ are faster, yet since the temperature at this point in time (same as \S\ref{subsubsec:87chart1037}) is lower, the the capture rates on lighter isotopes than $\textrm{Mn}$ are slower.

\begin{figure}[tbph]
\centering
 \plotone{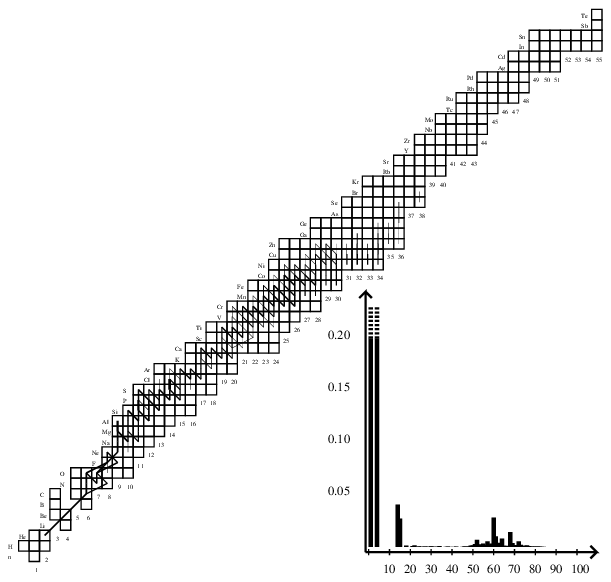} 
 \caption{Above ignition: $T=4.44\cdot10^8\textrm{K}$, $\rho=3.46\cdot10^5\textrm{g}/\textrm{cm}^3$, $X=0.398$, $Y=0.402$, $t=-11.091\textrm{s}$. (see end of \S\ref{sec:flow} for an explanation of the diagram).}\label{fig:90chart1044} 
\end{figure}

\subsubsection{Fig.~\ref{fig:90chart1044}: $T=4.44\cdot10^8\textrm{K}$, $\rho=3.46\cdot10^5\textrm{g}/\textrm{cm}^3$, $X=0.398$, $Y=0.402$, $t=-11.091\textrm{s}$}\label{subsubsec:90chart1044}

${}^{14}\textrm{O}(\alpha,p){}^{17}\textrm{F}$ is now as strong as ${}^{15}\textrm{O}(\alpha,\gamma){}^{19}\textrm{Ne}$. The flow through the ${}^{31}\textrm{Cl}(\beta^+,\nu)$ $(T_{1/2}=0.270\textrm{s})$ waiting point is currently approximately equal to the flow through the ${}^{30}\textrm{S}$ $(T_{1/2}=1.08\textrm{s})$ waiting point, but the latter will quickly become dominant as rising temperatures prevent the formation of ${}^{31}\textrm{Cl}$ due photodisintegration. 

In the $\textrm{Zn}$ region, the flow stops at the long lived ${}^{59}\textrm{Cu}$ $(T_{1/2}=84.7\textrm{s})$ and ${}^{60}\textrm{Zn}$ $(T_{1/2}=4.29\textrm{m})$, but the temperature is not yet sufficiently high for proton captures to established a reaction flow to heaver nuclei, nor has sufficient time passed to allow a substantial amount of material to decay through these two nuclei. 
It is interesting to note that processing to heaver material either depends on the temperature becoming sufficiently high for the ${}^{60}\textrm{Zn}(p,\gamma)(\gamma,p)$ ${}^{61}\textrm{Ga}$ equilibrium to allow $(p,\gamma)$-reactions on ${}^{61}\textrm{Ga}$ or the temperature remaining sufficiently low for the flow to decay through the faster ${}^{59}\textrm{Cu}(\beta^+,\nu)$ ${}^{59}\textrm{Ni}$-reaction. 

\begin{figure}[tbph]
\centering
 \plotone{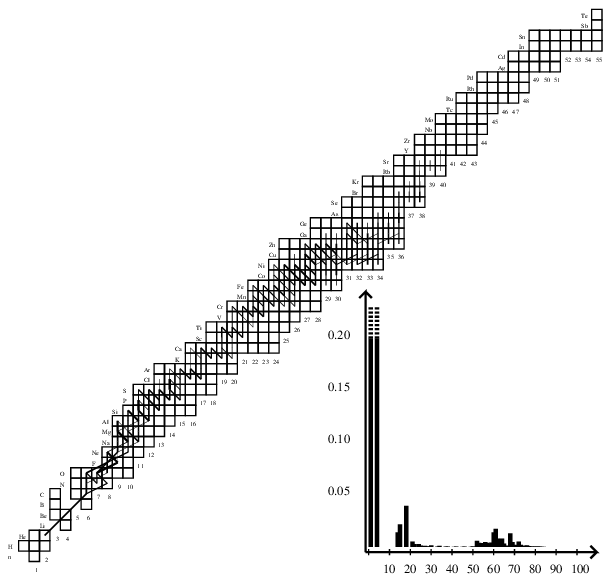} 
 \caption{Above ignition: $T=5.75\cdot10^8\textrm{K}$, $\rho=2.19\cdot10^5\textrm{g}/\textrm{cm}^3$, $X=0.393$, $Y=0.387$, $t=-10.418\textrm{s}$. (see end of \S\ref{sec:flow} for an explanation of the diagram).}\label{fig:90chart1101} 
\end{figure}

\subsubsection{Fig.~\ref{fig:90chart1101}: $T=5.75\cdot10^8\textrm{K}$, $\rho=2.19\cdot10^5\textrm{g}/\textrm{cm}^3$, $X=0.393$, $Y=0.387$, $t=-10.418\textrm{s}$}\label{subsubsec:90chart1101}
At this point the ${}^{18}\textrm{Ne}(\alpha,p)$ ${}^{21}\textrm{Na}$-reaction activates making it possible to move into the $rp$-process via ${}^{12}\textrm{C}(p,\gamma)$ ${}^{13}\textrm{N}(p,\gamma)$ ${}^{14}\textrm{O}(\alpha,p)$ ${}^{17}\textrm{F}(p,\gamma)$ ${}^{18}\textrm{Ne}(\alpha,p)$ ${}^{21}\textrm{Na}$ and so forth instead of waiting for the $T_{1/2}=1.67\textrm{s}$ beta-decay of ${}^{18}\textrm{Ne}$. The ${}^{18}\textrm{Ne}(\alpha,p)$ ${}^{21}\textrm{Na}$-reaction  is thus especially important as most of the energy release in the atmosphere originate in the $rp$-process at lower depths/higher hydrogen concentrations as seen in Fig.~\ref{fig:kippenhahn}.

However, presently $90\%$ of the the flow through the lighter isotopes stops at the ${}^{30}\textrm{S}$ $(T_{1/2}=1.09\textrm{s})$ waiting point with only a small flux following from its decay. This causes a temporary dip in the energy production, though the higher temperature ensures a flow close to the dripline from Ca to Ni. This flow is, however, slowed down at the $N=28$ isotones due to the long half-lives of ${}^{55}\textrm{Co}$ $(T_{1/2}=10.3\textrm{h})$ and ${}^{56}\textrm{Ni}$ $(T_{1/2}=24.9\textrm{h})$ which effectively prevents any beta decays of these isotopes. It is also interesting to note the $(p,\alpha)$-reactions on the heavier Cu isotopes going back to Ni while releasing hydrogen.

At this point there is a weak flow out of ${}^{59}\textrm{Cu}$ which allows additional proton captures viz. ${}^{59}\textrm{Ni}(p,\gamma)$${}^{60}\textrm{Cu}(p,\gamma)$ ${}^{61}\textrm{Zn}(p,\gamma)$ ${}^{62}\textrm{Ga}$ which can either decay or capture an additional proton to ${}^{63}\textrm{Ge}$ which then decays. If ${}^{62}\textrm{Ga}$ decays it captures two additional protons and goes to ${}^{64}\textrm{Ge}$. It is also possible to reach ${}^{65}\textrm{Ge}$ if ${}^{63}\textrm{Ga}$ decays. 

Heavier isotopes are generated through proton captures on ashes from the previous burst. None of these are beta decaying though. Ë

\begin{figure}[tbph]
\centering
\plotone{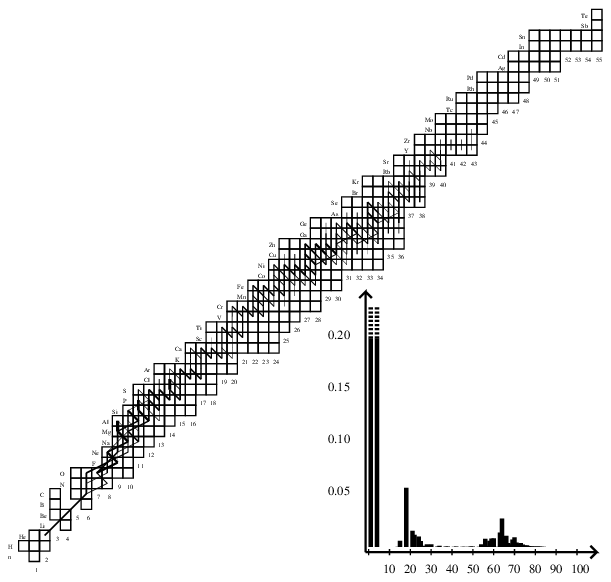} 
 \caption{Above ignition: $T=6.97\cdot10^8\textrm{K}$, $\rho=2.51\cdot10^5\textrm{g}/\textrm{cm}^3$, $X=0.381$, $Y=0.372$, $t=-9.994\textrm{s}$. (see end of \S\ref{sec:flow} for an explanation of the diagram).}\label{fig:90chart1143} 
\end{figure}

\subsubsection{Fig.~\ref{fig:90chart1143}: $T=6.97\cdot10^8\textrm{K}$, $\rho=2.51\cdot10^5\textrm{g}/\textrm{cm}^3$, $X=0.381$, $Y=0.372$, $t=-9.994\textrm{s}$}\label{subsubsec:90chart1143}
A couple of seconds after its breakout, ${}^{14}\textrm{O}(\alpha,p)$ ${}^{17}\textrm{F}$, becomes so fast that any ${}^{14}\textrm{O}$ is immediately destroyed. Consequently ${}^{15}\textrm{O}$ is only created via the hot CNO bi-cycle. However, the bi-cycle will become void, because it is now sufficiently hot for alpha-particles to penetrate the Coulomb barrier of ${}^{18}\textrm{Ne}$, thus skipping its $T_{1/2}=1.67\textrm{s}$ $\beta^+$-decay. 

Additional $(\alpha,p)$-captures now happen on ${}^{21}\textrm{Mg}$ and ${}^{25}\textrm{Si}$. The latter circumvents the $(T_{1/2}=0.176\textrm{s})$ half-life of ${}^{25}\textrm{Si}$, thus shortening the characteristic reaction flow timescale in the $A=20$--$30$ region. 
The timescale is dominated by the ${}^{30}\textrm{S}(\beta^+,\nu)$ $(T_{1/2}=1.09\textrm{s})$ ${}^{30}\textrm{P}$ reaction since $98\%$ of the flow passes through this reaction.

Another reaction is ${}^{24}\textrm{Si}(\alpha,p)$ ${}^{27}\textrm{P}$ but that is not as significant since  ${}^{24}\textrm{Si}$ gets destroyed by photodisintegration. It is interesting to note a weak but present ${}^{18}\textrm{Ne}(\alpha,\gamma)$ ${}^{22}\textrm{Mg}$ which competes with the $(\alpha,p)$-process. 

The flow up to ${}^{58}\textrm{Ni}$ remains the same. The increased temperature and flow sets up NiCuZn, ZnGaGe and GeAsSe cycles on ${}^{58}\textrm{Ni}$, ${}^{60}\textrm{Zn}$, and ${}^{66}\textrm{Ge}$. 

We note that there is still no flow out of ${}^{64}\textrm{Ge}(p,\gamma)$ ${}^{65}\textrm{As}$ as ${}^{65}\textrm{As}$ is weakly proton bound and 2p capture \citep{Schatz98} is not effective. Therefore the reaction flow proceeds via the slow beta decay ${}^{64}\textrm{Ge}(\beta^+,\nu)$ $(T_{1/2}=84.9\textrm{s})$ ${}^{64}\textrm{Ga}$ or the lighter ${}^{63}\textrm{Ga}(\beta^+,\nu)$ $( T_{1/2}=26.6\textrm{s})$ ${}^{63}\textrm{Zn}$. 

Reaching the $N=33$ isotones, the flow reaches ${}^{67}\textrm{Se}$ $( T_{1/2}=0.060\textrm{s})$ and ${}^{68}\textrm{Se}$ $( T_{1/2}=35.5\textrm{s})$. Further progress either depends on another 2p-reaction \citep{Schatz98} or ${}^{68}\textrm{Se}$ or ${}^{67}\textrm{As}$ decaying.

As similar challenge is posed by ${}^{72}\textrm{Kr}$ $(T_{1/2}=17.2\textrm{s})$, ${}^{76}\textrm{Sr}$ $(T_{1/2}=8.9\textrm{s})$ and ${}^{80}\textrm{Zr}$ $(T_{1/2}=3.9\textrm{s})$. Presently the flow has not moved farther although protons have started captures on heavier isotopes of the ashes of the previous burst. 

\begin{figure}[tbph]
\centering
\plotone{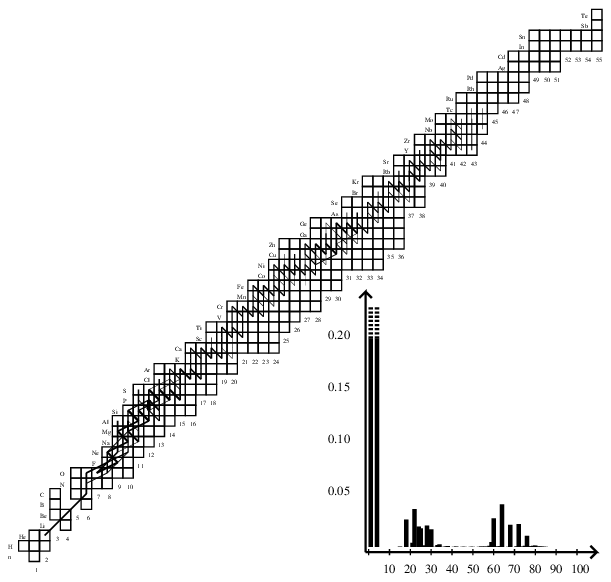} 
 \caption{Above ignition: $T=8.34\cdot10^8\textrm{K}$, $\rho=2.15\cdot10^5\textrm{g}/\textrm{cm}^3$, $X=0.358$, $Y=0.346$, $t=-9.097\textrm{s}$. (see end of \S\ref{sec:flow} for an explanation of the diagram).}\label{fig:90chart1197} 
\end{figure}

\subsubsection{Fig.~\ref{fig:90chart1197}: $T=8.34\cdot10^8\textrm{K}$, $\rho=2.15\cdot10^5\textrm{g}/\textrm{cm}^3$, $X=0.358$, $Y=0.346$, $t=-9.097\textrm{s}$}\label{subsubsec:90chart1197}
At this time the temperature is sufficiently high for photodisintegration of ${}^{27}\textrm{S}$ to prevent the shortcut, which was previously established between ${}^{25}\textrm{Si}$ and ${}^{27}\textrm{P}$.

However, at the same temperature the ${}^{21}\textrm{Mg}(\alpha,p)$ ${}^{24}\textrm{Al}$ and the ${}^{22}\textrm{Mg}(\alpha,p)$ ${}^{25}\textrm{Al}$ reactions become significant. In addition, ${}^{24}\textrm{Si}(\alpha,p)$ ${}^{27}\textrm{P}$, ${}^{25}\textrm{Si}(\alpha,p)$ ${}^{28}\textrm{P}$, and ${}^{26}\textrm{Si}(\alpha,p)$ ${}^{29}\textrm{P}$ become significant. 
Circumventing the ${}^{30}\textrm{S}(\beta^+,\nu)$ $(T_{1/2}=1.09\textrm{s})$ ${}^{30}\textrm{P}$ becomes possible through the ${}^{29}\textrm{S}(\alpha,p)$ ${}^{32}\textrm{Cl}$-reaction which at this point is not as strong as the beta decay.

 ${}^{38}\textrm{Ca}$ starts photodisintegration, but since ${}^{39}\textrm{Sc}$ is proton unbound, the flow must await a $(T_{1/2}=0.04\textrm{s})$-decay, since this reaction is a bottleneck.

Heavier isotopes with $N>32$ continue to ${}^{84}\textrm{Mo}$ $(T_{1/2}=3.6\textrm{s})$. Some hydrogen is capturing on $N=43$ and $N=44$ isotones of the ashes from the previous burst.

\begin{figure}[tbph]
\centering
\plotone{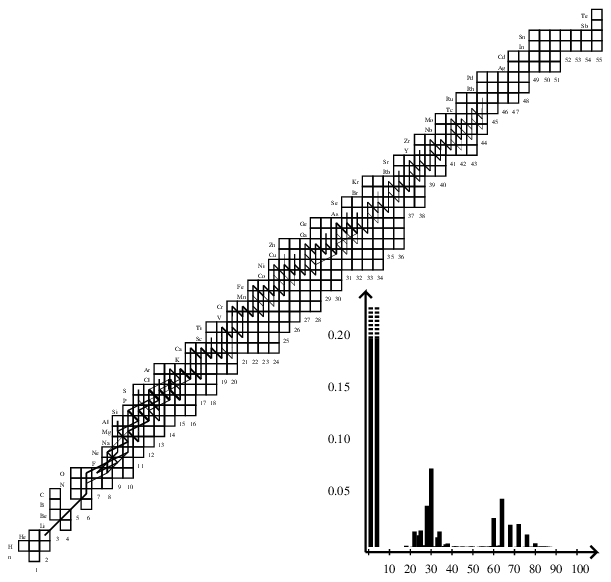} 
 \caption{Above ignition: $T=8.96\cdot10^8\textrm{K}$, $\rho=2.07\cdot10^5\textrm{g}/\textrm{cm}^3$, $X=0.327$, $Y=0.326$, $t=-8.075\textrm{s}$. (see end of \S\ref{sec:flow} for an explanation of the diagram).}\label{fig:90chart1225} 
\end{figure}

\subsubsection{Fig.~\ref{fig:90chart1225}: $T=8.96\cdot10^8\textrm{K}$, $\rho=2.07\cdot10^5\textrm{g}/\textrm{cm}^3$, $X=0.327$, $Y=0.326$, $t=-8.075\textrm{s}$}\label{subsubsec:90chart1225}

At this time, the ${}^{28}\textrm{S}(\alpha,p)$ ${}^{31}\textrm{Cl}$, ${}^{29}\textrm{S}(\alpha,p)$ ${}^{32}\textrm{Cl}$, and ${}^{30}\textrm{S}(\alpha,p)$ ${}^{33}\textrm{Cl}$ are all active. The latter now competes directly with the ${}^{30}\textrm{S}$ beta decay. This competition is especially important at lower accretion rates \citep{Fisker04b}. 

Heavier isotopes continue to ${}^{88}\textrm{Ru}$ $(T_{1/2}=1.1\textrm{s})$. There are no significant captures on heavier isotopes. 

\begin{figure}[tbph]
\centering
\plotone{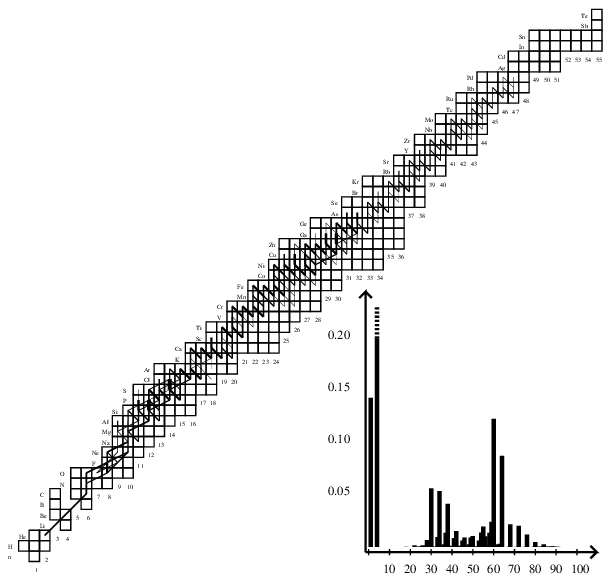} 
 \caption{Above ignition: $T=9.93\cdot10^8\textrm{K}$, $\rho=2.43\cdot10^5\textrm{g}/\textrm{cm}^3$, $X=0.143$, $Y=0.234$, $t=-3.013\textrm{s}$. (see end of \S\ref{sec:flow} for an explanation of the diagram).}\label{fig:90chart1376} 
\end{figure}

\subsubsection{Fig.~\ref{fig:90chart1376}: $T=9.93\cdot10^8\textrm{K}$, $\rho=2.43\cdot10^5\textrm{g}/\textrm{cm}^3$, $X=0.143$, $Y=0.234$, $t=-3.013\textrm{s}$}\label{subsubsec:90chart1376}
This region has now reached its maximum temperature. The reaction flow-path is very similar to the flow in Fig.~\ref{fig:90chart1225}.

One notable difference is the ${}^{34}\textrm{Ar}(\alpha,p)$ ${}^{37}\textrm{K}$ reaction which is the last among the $(\alpha,p)$-reactions for the temperatures encountered in type I XRBs.

Additionally, heavier isotopes continue to ${}^{92}\textrm{Pd}$ and ${}^{93}\textrm{Pd}$. This effectively constitutes the end of the $rp$-process which is short of the prediction of \cite{Schatz01}. The reason is the  thermal and compositional inertia as well as the much lower peak temperature achieved by our model. If these are ignored, the flow does reach the SnSbTe cycle as shown by \cite{Woosley04}.

This flow structure is maintained until hydrogen runs out.
 
\begin{figure}[tbph]
\centering
\plotone{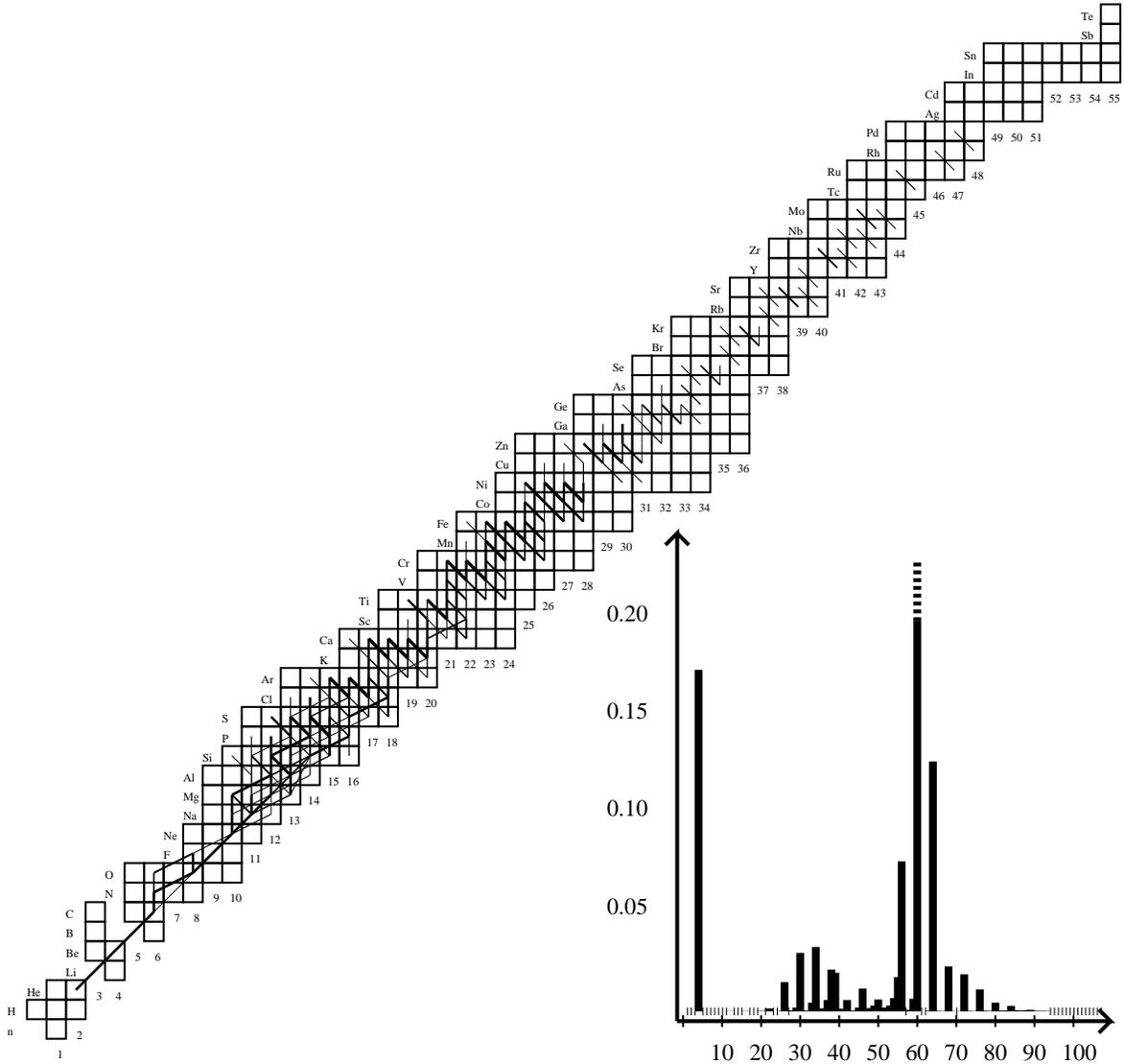} 
 \caption{Above ignition: $T=9.62\cdot10^8\textrm{K}$, $\rho=3.54\cdot10^5\textrm{g}/\textrm{cm}^3$, $X=2.5\times 10^{-5}$, $Y=0.175$, $X_{60}=0.346$, $t=1.476\textrm{s}$. (see end of \S\ref{sec:flow} for an explanation of the diagram).}\label{fig:90chart1483} 
\end{figure}

\subsubsection{Fig.~\ref{fig:90chart1483}: $T=9.62\cdot10^8\textrm{K}$, $\rho=3.54\cdot10^5\textrm{g}/\textrm{cm}^3$, $X=2.5\times 10^{-5}$, $Y=0.175$, $X_{60}=0.346$, $t=1.476\textrm{s}$}\label{subsubsec:90chart1483}
The last protons capture on the currently most abundant nuclei, namely the isotopes in the Ca-Ge region as the flow falls back towards the valley of stability as it is decaying along constant mass numbers. 

With helium still burning, the $(\alpha,p)$-process is still active along with previously mentioned $(\alpha,p)$reaction up to $A=36$. However, the ${}^{12}\textrm{C}(p,\gamma)$ ${}^{13}\textrm{N}(\alpha,p)$ ${}^{16}\textrm{O}$ path prevents much formation of ${}^{17}\textrm{F}$. This means that the $(\alpha,p)$-process starts on ${}^{21}\textrm{Na}$ which is reached from ${}^{16}\textrm{O}(\alpha,\gamma)$ ${}^{20}\textrm{Ne}(p,\gamma)$ ${}^{21}\textrm{Na}$.

\begin{figure}[tbph]
\centering
\plotone{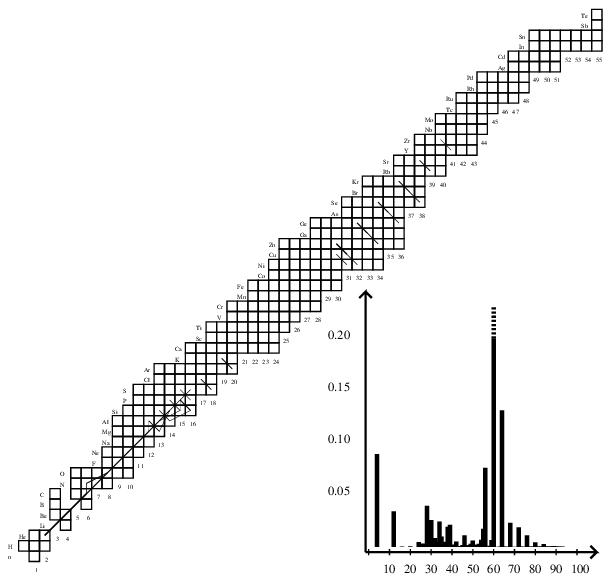} 
 \caption{Above ignition: $T=6.97\cdot10^8\textrm{K}$, $\rho=5.10\cdot10^5\textrm{g}/\textrm{cm}^3$, $X=1.0\times10^{-11}$, $Y=0.089$, $X_{60}=0.347$, $t=28.788\textrm{s}$. (see end of \S\ref{sec:flow} for an explanation of the diagram).}\label{fig:90chart1875} 
\end{figure}

\subsubsection{Fig.~\ref{fig:90chart1875}: $T=6.97\cdot10^8\textrm{K}$, $\rho=5.10\cdot10^5\textrm{g}/\textrm{cm}^3$, $X=1.0\times10^{-11}$, $Y=0.089$, $X_{60}=0.347$, $t=28.788\textrm{s}$}\label{subsubsec:90chart1875}
Half a minute after the burst the $rp$-process no longer operating and the temperature has decreased $30\%$ from the maximum temperature. 

An $(\alpha,\gamma)$-chain connects ${}^{16}\textrm{O}$ to ${}^{32}\textrm{S}$ which eliminates most of the ${}^{12}\textrm{C}$ via ${}^{12}\textrm{C}(p,\gamma)$ ${}^{13}\textrm{N}(\alpha,p)$ ${}^{16}\textrm{O}$. Meanwhile, heavier isotopes follow constant $A$ decay-chains back to the valley of stability. 

\subsection{Convective region}
The size of the convective region is shown in Fig.~\ref{fig:rhot}, which shows a trace of the burst conditions for different depths (pressures) during a complete revolution of the limit cycle. 
Note that the convective zone only exist during the phase where the temperature rises (the cycle revolves clockwise). 
The figure shows that the convective zone does not reach the top of our model for this burst, but stays in a narrow region between $y=5.7\times 10^6 \textrm{g}/\textrm{cm}^2$ and $y=5.2\times 10^7 \textrm{g}/\textrm{cm}^2$. Additionally, this burst does not reach super-Eddington luminosities, so no ashes will be ejected by a radiatively driven wind; something which is possible in helium-ignited bursts \citep{Weinberg06}.

The quantitative analysis of the turbulent convective burning is complicated by the mixing of matter between convective zones, which occurs as soon as and as long as a slightly superadiabatic temperature gradient is established.
However, the convective timescale, $\tau_{con.}\equiv\Lambda/v_{edd.}\sim10^{-6}\textrm{--}10^{-5}\textrm{s}\ll\tau_{rp}$, is generally faster than the typical timescale of the $rp$-process, so the explosive burning will have almost the same composition throughout the entire convective zone (see the convective model of \cite{Rembges99} which assumes identical composition throughout the convective zone for comparison) although burning happens at different temperatures and densities at the top and bottom of the convective zone respectively. 
Furthermore, turbulent convective burning does not happen above temperatures of $7\times 10^8\,\textrm{K}$, so the $(\alpha,p)$-process, which has a much shorter timescale, does not become active. 
Therefore this region could be computed by models with a simplified description of the compositional evolution but a more complex (2D) hydrodynamical implementation.
The following analysis of this region concentrates on the bottom of the convective region, because it is hotter and denser, and therefore the reactions proceed faster here.

\begin{figure}[tbph]
\centering
\plotone{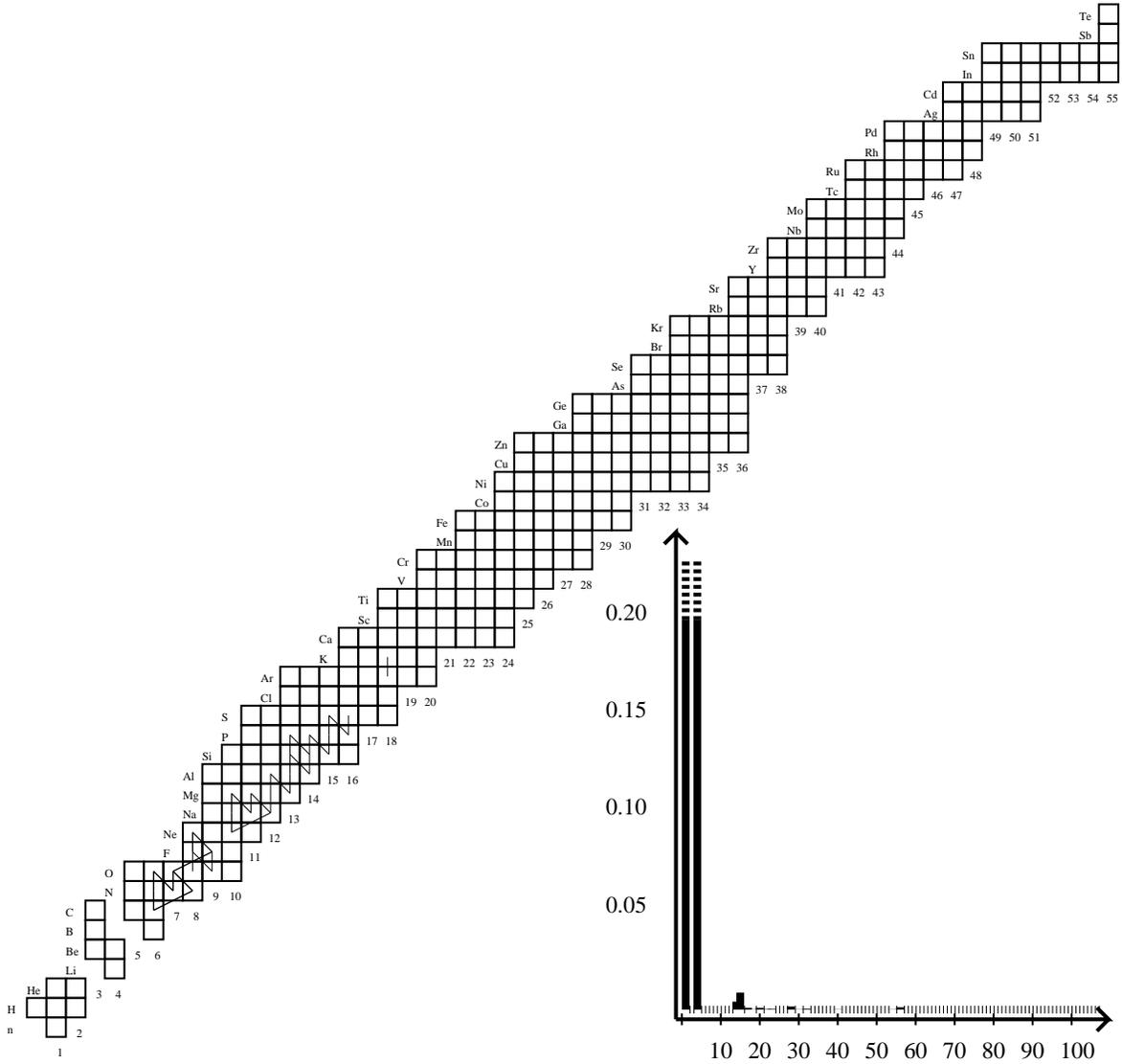} 
 \caption{Convection region: $T=3.92\cdot10^8\textrm{K}$, $\rho=1.161\cdot10^4\textrm{g}/\textrm{cm}^3$, $X=0.610$, $Y=0.363$, $t=-10.229\textrm{s}$. (see end of \S\ref{sec:flow} for an explanation of the diagram).}\label{fig:104chart1115} 
\end{figure}

\subsubsection{Fig.~\ref{fig:104chart1115}: $T=3.92\cdot10^8\textrm{K}$, $\rho=1.161\cdot10^4\textrm{g}/\textrm{cm}^3$, $X=0.610$, $Y=0.363$, $t=-10.229\textrm{s}$}\label{subsubsec:104chart1115}
The convective period during this burst lasts about $2.2\,\textrm{s}$ during which fresh unburned matter from the colder top of the convective zone mixes into the warmer bottom and back again. This means that temperature dependent particle-captures are effectively weaker, whereas the weak decays remain unaltered. 

Due to the short duration, the region attains a maximum temperature of $\sim 0.7\cdot10^9\textrm{K} \textrm{K}$ while it is convecting ($\sim 0.9\cdot10^9\textrm{K}$ at the peak), so the high temperature short cuts like $(\alpha,p)$-reactions or the upper leg of $rp$-process bifurcations available to the deeper layers and described in the previous sections never come into play; instead the initial reaction flow is mostly represented by the flow chart of Fig.~\ref{fig:104chart1115}.

At this point, the second hot CNO bi-cycle, ${}^{14}\textrm{O}(\alpha,p)$ ${}^{17}\textrm{F}(p,\gamma)$ ${}^{18}\textrm{Ne}(\beta^+,\nu)$ $(T_{1/2}=1.67\textrm{s})$ ${}^{18}\textrm{F}(p,\alpha){}^{15}\textrm{O}$ already dominates the first hot CNO bi-cycle, ${}^{19}\textrm{Ne}(\beta^+,\nu)$ $(T_{1/2}=17.2\textrm{s})$ ${}^{19}\textrm{F}(p,\alpha)$ ${}^{16}\textrm{O}(p,\gamma)$  ${}^{17}\textrm{F}(p,\gamma)$ ${}^{18}\textrm{Ne}(\beta^+,\nu)$ $(T_{1/2}=1.67\textrm{s})$ ${}^{18}\textrm{F}(p,\alpha){}^{15}\textrm{O}$, 
as the temperature is high enough for the ${}^{19}\textrm{Ne}(p,\gamma)$ ${}^{20}\textrm{Na}$-reaction to dominate the slow ${}^{19}\textrm{Ne}(\beta^+,\nu)$ $(T_{1/2}=17.2\textrm{s})$ ${}^{19}\textrm{F}$. From this point ${}^{20}\textrm{Na}$ captures another proton so ${}^{20}\textrm{Na}(p,\gamma)$ ${}^{21}\textrm{Mg}(\beta^+,\nu)$ $(T_{1/2}=0.124\textrm{s})$ ${}^{21}\textrm{Na}(p,\gamma)$ ${}^{22}\textrm{Mg}$. 

Here the flow bifurcates to either ${}^{22}\textrm{Mg}(\beta^+,\nu)$ $(T_{1/2}=3.46\textrm{s})$ ${}^{22}\textrm{Na}(p,\gamma)$ ${}^{23}\textrm{Mg}(p,\gamma)$ ${}^{24}\textrm{Al}$ or ${}^{22}\textrm{Mg}(p,\gamma)$ ${}^{23}\textrm{Al}(p,\gamma)$ ${}^{24}\textrm{Si}(\beta^+,\nu)$ $( T_{1/2}=0.191\textrm{s})$ ${}^{24}\textrm{Al}$. A similar bifurcation exists at ${}^{25}\textrm{Si}(p,\gamma)$ ${}^{26}\textrm{P}(p,\gamma)$ ${}^{27}\textrm{S}(\beta^+,\nu)$ $( T_{1/2}=0.021\textrm{s})$ ${}^{27}\textrm{P}$ competing with ${}^{25}\textrm{Si}(\beta^+,\nu)$ $( T_{1/2}=0.189\textrm{s})$ ${}^{25}\textrm{Al}(p,\gamma)$ ${}^{26}\textrm{Si}(p,\gamma)$ ${}^{27}\textrm{P}$.

The ${}^{30}\textrm{S}$ waiting point still acts as a bottleneck with a small leak via ${}^{30}\textrm{S}(p,\gamma)$ ${}^{31}\textrm{Cl}(\beta^+,\nu)$ $( T_{1/2}=0.272\textrm{s})$ ${}^{30}\textrm{S}$. However, this is quickly reduced by photodisintegration as before.

This flow passes through the ${}^{34}\textrm{Ar}(\beta^+,\nu)$ $( T_{1/2}=0.824\textrm{s})$ bottleneck and on through ${}^{37}\textrm{Ca}(\beta^+,\nu)$ $( T_{1/2}=0.155\textrm{s})$ ${}^{37}\textrm{K}(p,\gamma)$ ${}^{38}\textrm{Ca}(\beta^+,\nu)$ $( T_{1/2}=0.423\textrm{s})$ ${}^{38}\textrm{K}(p,\gamma)$ ${}^{39}\textrm{Ca}(\beta^+,\nu)$ $( T_{1/2}=0.808\textrm{s})$ ${}^{39}\textrm{K}(p,\gamma)$ ${}^{40}\textrm{Ca}$ into the $pf$-shell isotopes.

It is interesting to notice that the $rp$-process is already active in the ${}^{40}\textrm{Ca}$--${}^{52}\textrm{Fe}$ region. This is because the double-magic ${}^{40}\textrm{Ca}$ is the natural end-point of the reactions during the minute long run up to the burst ignition where a small reaction flow is already present. 

\begin{figure}[tbph]
\centering
\plotone{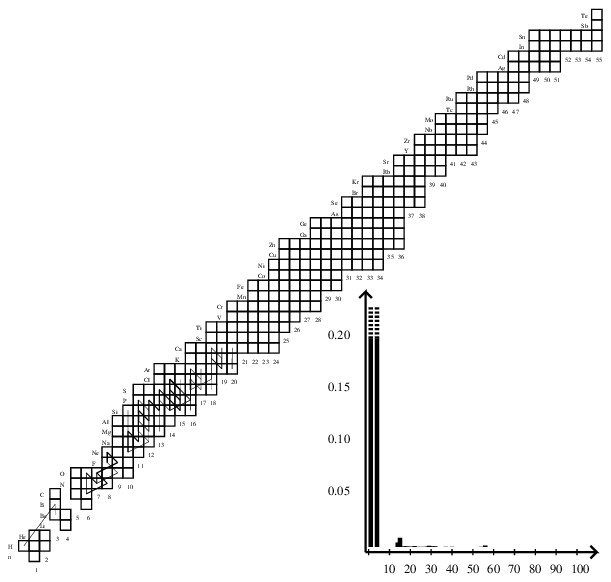} 
 \caption{Convection region: $T=6.90\cdot10^8\textrm{K}$, $\rho=9.35\cdot10^4\textrm{g}/\textrm{cm}^3$, $X=0.646$, $Y=0.328$, $t=-8.600\textrm{s}$. (see end of \S\ref{sec:flow} for an explanation of the diagram).}\label{fig:104chart1214} 
\end{figure}

\subsubsection{Fig.~\ref{fig:104chart1214}: $T=6.90\cdot10^8\textrm{K}$, $\rho=9.35\cdot10^4\textrm{g}/\textrm{cm}^3$, $X=0.646$, $Y=0.328$, $t=-8.600\textrm{s}$}\label{subsubsec:104chart1214}
Fig.~\ref{fig:104chart1214} shows the maximum temperature during the convective phase occuring concurrently with the maximum temperature in the ignition region, which drives the superadiabatic temperature gradient responsible for the convective turnover. Higher temperatures are reached at this depth, but after this time, the region is no longer convective.

At the end of the convective phase, the $(\alpha,p)$-process at the bottom of the convective region extends to ${}^{26}\textrm{Si}$. Following that, the flow to heaver isotopes is impacted by the ${}^{30}\textrm{S}$ $(T_{1/2}=1.10\textrm{s})$ waiting point. This is significant because the entire convective phase only lasts two half lives of ${}^{30}\textrm{S}$. Similarly, there are bottlenecks at ${}^{59}\textrm{Cu}$ $(T_{1/2}=91.9\textrm{s})$ and ${}^{60}\textrm{Zn}$ $(T_{1/2}=5.28\textrm{m})$ which require a ${}^{60}\textrm{Zn}(p,\gamma)$ ${}^{61}\textrm{Ga}$ breakout that does not happen at these temperatures during the short convection phase.

\subsection{Surface region}
\begin{figure}[tbph]
\centering
\plotone{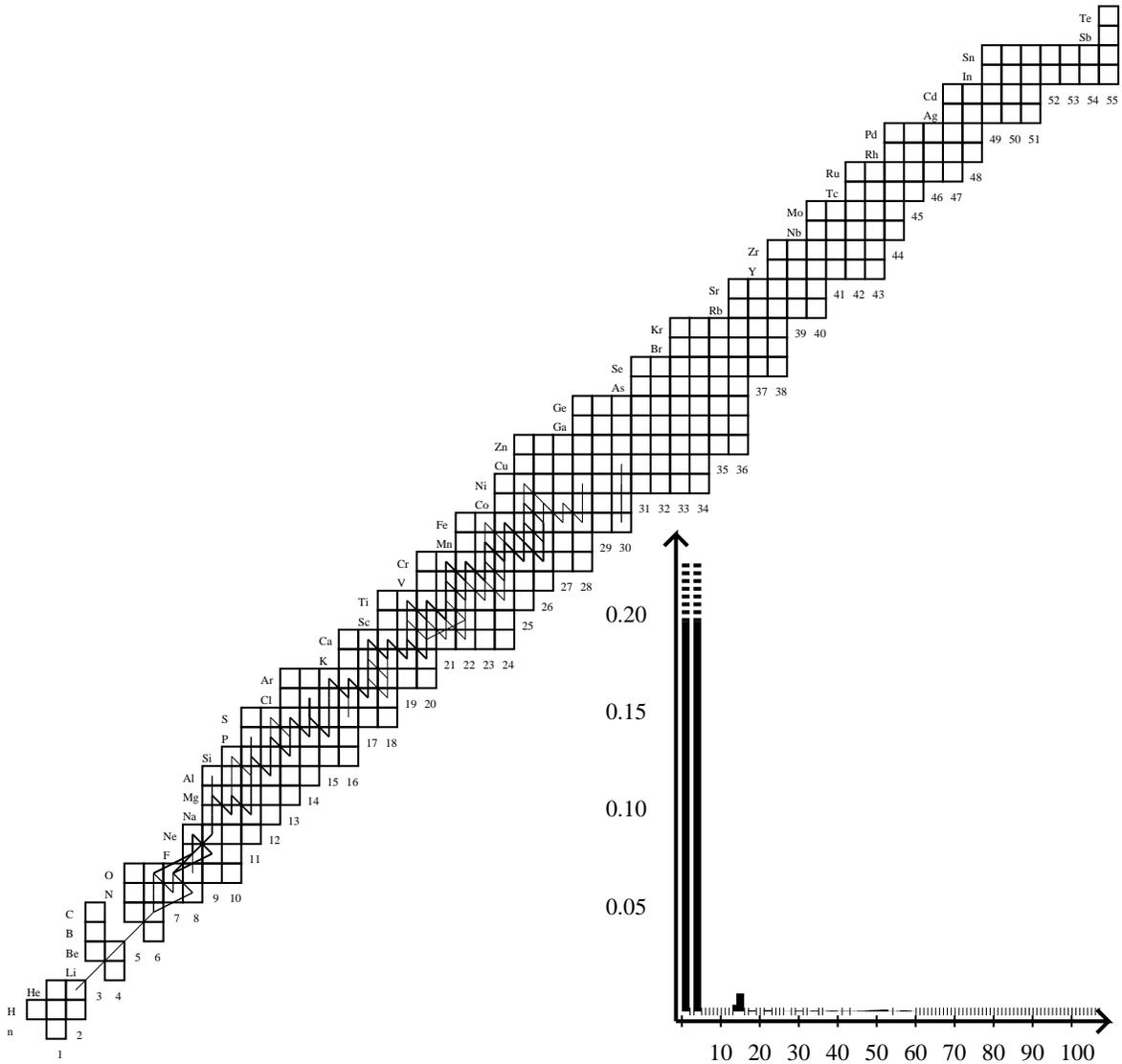} 
 \caption{Surface: $T=5.31\cdot10^8\textrm{K}$, $\rho=8.75\cdot10^3\textrm{g}/\textrm{cm}^3$, $X=0.697$, $Y=0.281$, $t=-0.209\textrm{s}$. (see end of \S\ref{sec:flow} for an explanation of the diagram).}\label{fig:128chart1419} 
\end{figure}

In H/He-ignited XRBs the convective region does not extend to the top of our model. This means that if the convective model does not severely underestimate the convective strength then heavier ashes are not brought to the surface, 
Since the matter at the top of our model is extremely opaque with mean free photon paths of $\sim 10^{-4}\textrm{cm}$, the photons are in local thermal equilibrium (LTE) and exhibit a black body spectrum with no lines. Comparison between the results of this section with observations therefore require this model to be coupled with a radiative transport code (see \cite{Weinberg06}).
\subsubsection{Fig.~\ref{fig:128chart1419}: $T=5.31\cdot10^8\textrm{K}$, $\rho=8.75\cdot10^3\textrm{g}/\textrm{cm}^3$, $X=0.697$, $Y=0.281$, $t=-0.209\textrm{s}$}\label{subsubsec:128chart1419}

The extent of the reaction flow at the maximum temperature is shown in Fig.~\ref{fig:128chart1419} and ends at ${}^{56}\textrm{Ni}$. 
This region is limited by $T<5.3\cdot10^8\textrm{K}$ and the initial reactions are characterized by proton captures on the accreted heavy elements, which may have been destroyed by the surface impact \citep{Bildsten92}.

\subsection{Ocean (ashes)}\label{subsec:ocean}
The inner parts of the neutron star acts as a buffer absorbing heat from the burst. However, for this accretion rate it is radiated outwards again after the burst, therefore it does not heat the crust (\cite{Fujimoto84}). 
\begin{figure}[tbph]
\centering
\plotone{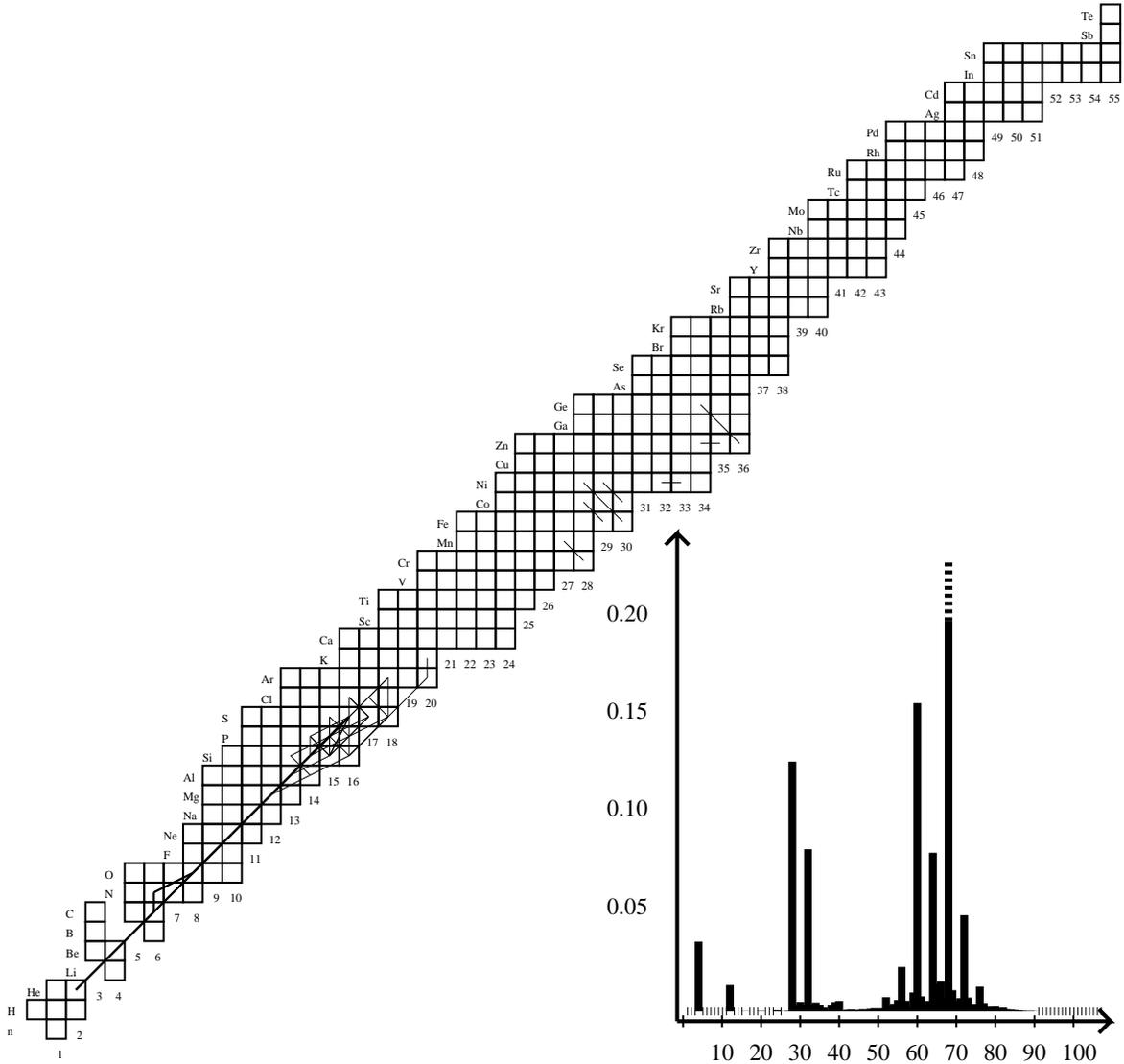} 
 \caption{Ocean: $T=9.55\cdot10^8\textrm{K}$, $\rho=6.06\cdot10^5\textrm{g}/\textrm{cm}^3$, $X=5.3\cdot 10^{-11}$, $Y=0.036$, $X_{68}=0.22$, $t=-0.209\textrm{s}$. (see end of \S\ref{sec:flow} for an explanation of the diagram).}\label{fig:81chart1481} 
\end{figure}

The early reaction flow which is caused by conductive heating in a hydrogen depleted environment is similar to the reaction flow in Fig.~\ref{fig:87chart1375}.
Later it is characterized by residual helium, which has been advected down from the previous burst, capturing on alpha-chain nuclei extending to ${}^{36}\textrm{Ar}$ as shown in Fig.~\ref{fig:81chart1481}. Note that here we also have ${}^{12}\textrm{C}(p,\gamma)$ ${}^{13}\textrm{N}(\alpha,p)$ ${}^{16}\textrm{O}$ being much stronger than the direct ${}^{12}\textrm{C}(\alpha,\gamma)$ ${}^{16}\textrm{O}$-reaction. Here the protons are supplied by many weak $(\alpha,p)$-reactions on stable isotopes resulting from long lived $\beta^+$-decays in the sulfur region in matter that has advected downwards from above.

\section{Conclusion}
Important in all the regions are the hot-CNO cycle and its respective breakout reactions, the $(\alpha,p)$-process, as well as $(p,\gamma)(\gamma,p)$-equilibria and waiting points of the $rp$-process. These are now discussed. 

\subsection{Hot CNO cycles}
There are essentially three hot CNO cycles, specifically, the first hot CNO cycle: ${}^{12}\textrm{C}(p,\gamma)$ ${}^{13}\textrm{N}(p,\gamma)$ ${}^{14}\textrm{O}(\beta^+,\nu)$ ${}^{14}\textrm{N}(p,\gamma)$ ${}^{15}\textrm{O}(\beta^+,\nu)$ ${}^{15}\textrm{N}(p,\alpha)$ ${}^{12}\textrm{C}$, the second hot CNO cycle: ${}^{14}\textrm{O}(\alpha,p)$ ${}^{17}\textrm{F}(p,\gamma)$ ${}^{18}\textrm{Ne}(\beta^+,\nu)$ ${}^{18}\textrm{F}(p,\alpha){}^{15}\textrm{O}$ and third hot CNO cycle ${}^{15}\textrm{O}(\alpha,\gamma)$ ${}^{19}\textrm{Ne}(\beta^+,\nu)$ ${}^{19}\textrm{F}(p,\alpha)$ ${}^{16}\textrm{O}(p,\gamma)$  ${}^{17}\textrm{F}(p,\gamma)$ ${}^{18}\textrm{Ne}(\beta^+,\nu)$ ${}^{18}\textrm{F}(p,\alpha){}^{15}\textrm{O}$. 

In order to activate the second hot CNO cycle, via ${}^{14}\textrm{O}(\alpha,p)$ ${}^{17}\textrm{F}$ and its breakout via ${}^{17}\textrm{F}(p,\gamma)$ ${}^{18}\textrm{Ne}(\alpha,p)$ ${}^{21}\textrm{Na}$, the third hot CNO cycle must activate hundreds of second prior to the runaway and achieve a breakout via ${}^{19}\textrm{Ne}(p,\gamma)$ ${}^{20}\textrm{Na}$. If the ${}^{15}\textrm{O}(\alpha,\gamma)$ ${}^{19}\textrm{Ne}$-reaction is too weak, the third cycle never activates which means that the second cycle does not activate either and the thermonuclear runaway does not happen \citep{Fisker06}. These rates are therefore quite significant in connecting the hot CNO cycle and the $rp$-process. Additionally, the reaction flow, in particular the third hot CNO cycle of \cite{Cooper06}, prior to the runaway is important for the ignition composition as it influences the concentration of hydrogen and helium which is important to the thermonuclear instability.

\subsection{The $(\alpha,p)$-process}
The $(\alpha,p)$-process is important because it is a temperature dependent process unlike the $rp$-process that contains temperature-independent $\beta^+$-decays. The $(\alpha,p)$-process therefore influences the characteristic timescale of the reaction flow up to $A=36$ after which the Coulomb barrier becomes prohibitive. Furthermore, as shown in \cite{Fisker04b}, the $(\alpha,p)$-reactions in the $(\alpha,p)$-process lie on waiting points with ${}^{30}\textrm{S}$ being the most signifiance. Other potential waiting points are ${}^{34}\textrm{Ar}$ and ${}^{26}\textrm{Si}$. 

The most important $(\alpha,p)$-reactions for the XRB are therefore ${}^{26}\textrm{Si}(\alpha,p)$ ${}^{29}\textrm{P}$, ${}^{30}\textrm{S}(\alpha,p)$ ${}^{33}\textrm{Cl}$, and ${}^{34}\textrm{Ar}(\alpha,p)$ ${}^{37}\textrm{K}$. The ${}^{22}\textrm{Mg}(\alpha,p)$ ${}^{25}\textrm{Al}$-reaction is most likely not as important, since the flow moves through the ${}^{22}\textrm{Mg}$ waiting point via ${}^{22}\textrm{Mg}(p,\gamma)$ ${}^{23}\textrm{Al}$ before the $(\alpha,p)$-reaction becomes active. 

Other $(\alpha,p)$-reactions are less dominant since they operate at higher temperatures and on more proton-rich nuclei which are more susceptible to photodisintegration viz.~${}^{21}\textrm{Mg}(\alpha,p)$ ${}^{24}\textrm{Al}$, ${}^{24}\textrm{Si}(\alpha,p)$ ${}^{27}\textrm{P}$, ${}^{25}\textrm{Si}(\alpha,p)$ ${}^{28}\textrm{P}$, ${}^{28}\textrm{S}(\alpha,p)$ ${}^{31}\textrm{Cl}$, and ${}^{29}\textrm{S}(\alpha,p)$ ${}^{32}\textrm{Cl}$. The final $(\alpha,p)$-reaction is ${}^{13}\textrm{N}(\alpha,p)$ ${}^{16}\textrm{O}$ which in the event of hydrogen depletion is stronger than the ${}^{12}\textrm{C}(\alpha,\gamma)$ ${}^{16}\textrm{O}$ reaction (also see \cite{Weinberg06}). 

\subsection{The $rp$-process}
The $rp$-process evolution depends on the concentration of hydrogen and the peak temperature. The peak temperature is easily estimated as $P=a_{rad}T^4$ which assumes that the pressure is fully supported by radiation and that the dynamical pressure is negligible. This is a good assumption as the gravitational binding energy is a factor $\sim 20$--$50$ higher than nuclear energy release of the burst. This dependence means that if the pressure of the ignition point is inaccurately determined, the peak temperature may be off by $10\%$ or more which will significantly change the conclusions about the flow. Thermal and compositional inertia must be taken into account when considering the reaction flow. This was first shown by \cite{Woosley04} who started with a pure ${}^{56}\textrm{Fe}$ atmosphere which allowed accreted matter to reach deeper layers. As a result \cite{Woosley04} obtained the same results as the one-zone model of \cite{Schatz01} who based their ignition pressure and composition on analytical estimates. On the other hand, selfconsistently obtained bursts by \cite{Woosley04} match the results obtained by other selfconsistent models \citep{Rembges99, Fisker03, Fisker04b, Fisker05b,Fisker05a} as well as this paper. 

\subsubsection{$rp$-process waiting points}
Waiting points are isotopes from which further net reaction flows are (possibly temporarily) restricted due to either insufficiently high temperatures, insufficient capture particles, or the immediate photodisintegration due to a $(p,\gamma)(\gamma,p)$-equilibrium. Waiting points are easily identified by their temporary abundance spikes. If a substantial, say $20\%$ or more, part of the flow is backed up at a given isotope for a time comparable to the time scale of the XRB, it can significantly influence the shape of the observed luminosity curve (see \cite{Fisker04b}). 

During the early build up to the XRB and during the early phases of the ${}^{21}\textrm{Mg}(p,\gamma)(\gamma,p){}^{22}\textrm{Al}$-equilibrium, which depends on the $Q$-value of the proton capture reaction, means that ${}^{21}\textrm{Mg}$ must $\beta^+$-decay. The half life is short compared to the build-up phase which is on the order of hundreds of seconds. It is however comparable with the runaway time of the XRB. Therefore, the runaway depends on the ${}^{21}\textrm{Mg}(\alpha,p)$ ${}^{24}\textrm{Al}$ reaction. Similar waiting points can be found along the $(\alpha,p)$-process reaction path. They are ${}^{22}\textrm{Mg}$, ${}^{26}\textrm{Si}$, ${}^{30}\textrm{S}$, ${}^{34}\textrm{Ar}$, and ${}^{38}\textrm{Ca}$.
The dominant waiting point in this sequence depends on the extent of the $(\alpha,p)$-process which depends on the peak temperature. If the peak temperature is extremely high e.g. $T_{peak}>1.3\times 10^9\textrm{K}$ these waiting points are bypassed by the $(\alpha,p)$-process. For lower peak temperatures, these waiting point along with their associated $Q$-values and proton capture rates become important. 
However, our model has never reached peak temperatures above $\sim 1.3\times 10^9\textrm{K}$ for accretion rates greater than $\dot{M}=5\cdot 10^{16}\,\textrm{g}\,\textrm{s}^{-1}$ while accreting a solar composition \citep{Anders89} on a self-consistently attained atmosphere.
The ${}^{38}\textrm{Ca}$ waiting point might be circumvented by ${}^{38}\textrm{Ca}(p,\gamma)$ ${}^{39}\textrm{Sc}(p,\gamma)$ ${}^{40}\textrm{Ti}$ or ${}^{38}\textrm{Ca}(2p,\gamma)$ ${}^{40}\textrm{Ti}$.

Hot CNO-like cycles exist on well-bound isotopes such as ${}^{40}\textrm{Ca}$. This isotope is particularly interesting since the flow passes through it during the quiescent phase. The low-temperature ${}^{43}\textrm{Sc}(p,\gamma)$ ${}^{44}\textrm{Ti}$-reaction is therefore an important bottleneck as it determines the developing composition during the quiescent phase and thus the ignition conditions. 
The next bottleneck in the quiescent flow is ${}^{48}\textrm{Cr}(p,\gamma)$ ${}^{49}\textrm{Mn}$. During the burst (above $T\sim 5\cdot10^8\textrm{K}$) the flow through the Ca-Ni region goes through many $\beta^+$-decays and $(p,\gamma)$-reactions leaving no single determining reaction. Due to conservative scheme of numerical discretization of the model, it is possible to track minor variations in the luminosity due to individual rates. 

The Ni-Se region provides several waiting points. The first waiting points are ${}^{59}\textrm{Cu}$ and ${}^{60}\textrm{Zn}$. These half lives are on the order of the burst decay timescale and must be surpassed by proton captures in order for heavier isotopes to be produced. 
The latter is in $(p,\gamma)(\gamma,p)$-equilibrium and thus depends on the $Q$-value of ${}^{60}\textrm{Zn}(p,\gamma)$ ${}^{61}\textrm{Ga}$. There is a possible flow via ${}^{61}\textrm{Ga}(p,\gamma)$ ${}^{62}\textrm{Ge}$. A similar situation exists on ${}^{64}\textrm{Ge}$. Here ${}^{65}\textrm{As}$ is proton-unbound, so further flow depends on either a $2p$-capture \citep{Schatz98} or a slow $\beta^+$-decay. This is the reason why most of the flow stops at the $A=64$. The atmosphere cools before a substantial amount of matter can decay and be processed to heavier isotopes. Similar situations exist on ${}^{68}\textrm{Se}$, ${}^{72}\textrm{Kr}$, ${}^{76}\textrm{Sr}$, and ${}^{80}\textrm{Zr}$ where the corresponding ${}^{69}\textrm{Br}$, ${}^{73}\textrm{Rb}$, ${}^{77}\textrm{Y}$, and ${}^{81}\textrm{Nb}$ are also proton-unbound. These waiting points have also been identified by \cite{Schatz98} and by \cite{Woosley04}, who showed the significance of these decays by varying groups of electron capture and $\beta^+$-decay rates up and down by 1 order of magnitude thus testing the impact of the efficiency of the reaction flow progression through the waiting points on the burst lightcurve.

\subsection{Superbursts and convection}
We showed that the peak burst temperature is less than $\sim 1.3 \textrm{GK}$ and thus not as high as previously assumed for nuclear reaction studies. This means that $\textrm{Te}$ is not generated in quantity which corroborates previous multi-zone simulations of \cite{Fisker03,Fisker05a,Woosley04}. The average mass of the ashes is $\sim 64$ \citep{Fisker03,Fisker05a,Woosley04}. At the same time carbon is slowly destroyed by helium captures below the ignition zone and at the top of the ocean. This corroborates the findings of \cite{Woosley04} and it does not favor the parameter space requirements of current superburst theories \citep{Cumming01}. However, our model did not consider sedimentation effects which may change this conclusion \citep{Peng06}. 

We find that the convective region does not hit the top of our model for mixed hydrogen/helium (sub-Eddington) bursters. Therefore we predict that any spectral lines observed during such bursts are not from material that was burned at any significant depth. However, at lower accretion rates, the convective region does hit the top of our model for helium bursters (see \cite{Fisker05b}). 

\subsection{Summary}
The main result of our calculations is the identification of the nuclear reaction sequences that power type I X-ray bursts. In particular, we describe the complete nuclear reaction flow as a function of time and depth, including branchings and waiting points, as it evolves with realistic, rapidly changing temperatures and densities. Clearly, the reaction sequences are more complex than previously assumed based on the analysis of much simpler models. Our work is a necessary first step towards identifying the critical nuclear reaction rates in X-ray bursts that have the largest impact on observables such as light curves, or, indirectly, the composition of the ashes. One can then also begin to disentangle the effects of nuclear burning on the luminosity \citep{Fisker04b} from geometric effects such as the propagation of the burning front around the NS \citep{Spitkovsky02} to better explain the many different and somewhat inconsistent shapes of the burst luminosity profiles.

\acknowledgments
JLF and HS were supported by NSF-PFC grant PHY02--16783 through the Joint Institute of Nuclear Astrophysics\footnote{see \email{http://www.JINAweb.org}}. FKT and JLF acknowledge support from the Swiss NSF grant 20--068031.02. 


\end{document}